\documentclass[traditabstract]{aa} 
\usepackage{graphicx}
\usepackage{txfonts}
\usepackage{natbib}
\newcommand{\kms}{{\hbox {km\thinspace s$^{-1}$}}}
\newcommand{\Lsun}{{\hbox {L$_\odot$}}}
\newcommand{\Msun}{{\hbox {M$_\odot$}}}

\newcommand{\cmt}{{\hbox {cm$^{-3}$}}}
\newcommand{\cmd}{{\hbox {cm$^{-2}$}}}
\newcommand{\microns}{{\hbox {$\mu$m}}}
\newcommand{\hdo}{{\hbox {H$_{2}$O}}}
\newcommand{\nht}{{\hbox {NH$_{3}$}}}

\newcommand{\ohp}{{\hbox {OH$^+$}}}
\newcommand{\hdop}{{\hbox {H$_2$O$^+$}}}
\newcommand{\htop}{{\hbox {H$_{3}$O$^{+}$}}}

\newcommand{\tgas}{{\hbox {$T_{\mathrm{gas}}$}}}
\newcommand{\trot}{{\hbox {$T_{\mathrm{rot}}$}}}
\newcommand{\tform}{{\hbox {$T_{\mathrm{form}}$}}}
\newcommand{\tdust}{{\hbox {$T_{\mathrm{dust}}$}}}

\newcommand{\ccore}{{\hbox {$C_{\mathrm{core}}$}}}
\newcommand{\cwarm}{{\hbox {$C_{\mathrm{warm}}$}}}
\newcommand{\cext}{{\hbox {$C_{\mathrm{extended}}$}}}
\newcommand{\cwest}{{\hbox {$C_{\mathrm{west}}$}}}

\newcommand{\chalo}{{\hbox {$C_{\mathrm{halo}}$}}}

\def\t#1#2#3#4#5#6{{\hbox {$#1_{#2#3}\!\leftarrow\!#4_{#5#6}$}}}
\def\tohplus#1#2#3#4{{\hbox {$#1_{#2}\!\leftarrow\!#3_{#4}$}}}

\def\13co{$^{13}$CO}
\def\c18o{C$^{18}$O}

\begin{document}
   \title{Excited OH$^+$, H$_2$O$^+$, and H$_3$O$^+$ in NGC
     4418 and Arp 220\thanks{Herschel is an 
       ESA space observatory with science 
       instruments provided by European-led Principal Investigator consortia
       and with important participation from NASA.} }

   \author{E. Gonz\'alez-Alfonso \inst{1}, J. Fischer \inst{2}, 
S. Bruderer \inst{3}, H.~S.~P. M\"uller\inst{4}, 
J. Graci\'a-Carpio \inst{3}, E. Sturm \inst{3},  
D. Lutz \inst{3}, A. Poglitsch \inst{3},
H. Feuchtgruber \inst{3},  
S. Veilleux \inst{5,6}, A. Contursi \inst{3},
A. Sternberg \inst{7}, S. Hailey-Dunsheath \inst{8}, 
A. Verma \inst{9}, N. Christopher \inst{9},
R. Davies \inst{3},  
R. Genzel \inst{3}, L. Tacconi \inst{3}
}

   \institute{Universidad de Alcal\'a de Henares, 
Departamento de F\'{\i}sica y Matem\'aticas, Campus Universitario, E-28871
Alcal\'a de Henares, Madrid, Spain  
         \and
Naval Research Laboratory, Remote Sensing Division, 4555
     Overlook Ave SW, Washington, DC 20375, USA
         \and
Max-Planck-Institute for Extraterrestrial Physics (MPE), Giessenbachstra{\ss}e
1, 85748 Garching, Germany 
         \and
I.~Physikalisches Institut, Universit{\"a}t zu K{\"o}ln,
Z{\"u}lpicher Str. 77, 50937 K{\"o}ln, Germany
         \and
Department of Astronomy, University of Maryland, College Park, MD 20742, USA
         \and
Astroparticle Physics Laboratory, NASA Goddard Space
       Flight Center, Code 661, Greenbelt, MD 20771 USA
         \and
Sackler School of Physics \& Astronomy, Tel Aviv University, Ramat Aviv 69978,
Israel
         \and
California Institute of Technology, Mail Code 301-17, 1200 E. California
Blvd., Pasadena, CA 91125, USA
         \and
University of Oxford, Oxford Astrophysics, Denys Wilkinson Building, Keble
Road, Oxford, OX1 3RH, UK 
}


 
   \authorrunning{Gonz\'alez-Alfonso et al.}
   \titlerunning{Excited OH$^+$, H$_2$O$^+$, and H$_3$O$^+$ in NGC 4418 and Arp
     220}

  \abstract
   {We report on Herschel/PACS observations of absorption lines of OH$^+$,
     H$_2$O$^+$ and H$_3$O$^+$ in NGC~4418 and Arp~220. Excited lines of
     OH$^+$ and H$_2$O$^+$ with $E_{\mathrm{lower}}$ of at least 285 and
     $\sim200$ K, respectively, are detected in both sources,
     indicating radiative pumping and location in the high radiation density
     environment of the nuclear regions. Abundance ratios
     $\mathrm{OH^+/H_2O^+}$ of $1-2.5$ are estimated in the nuclei of both
     sources. The inferred OH$^+$ column and abundance relative to H nuclei
     are $(0.5-1)\times10^{16}$ \cmd\ and $\sim2\times10^{-8}$, respectively.
     Additionally, in Arp 220, an extended low excitation component around the
     nuclear region is found to have $\mathrm{OH^+/H_2O^+\sim5-10}$.   
     H$_3$O$^+$ is detected in both sources with
     $N$(H$_3$O$^+$)$\sim(0.5-2)\times10^{16}$ \cmd, and in Arp 220 the pure
     inversion, metastable lines indicate a high rotational temperature of
     $\sim500$ K, indicative of formation pumping and/or hot gas. Simple
     chemical models favor an ionization sequence dominated by $\mathrm{H^+
       \rightarrow O^+\rightarrow OH^+\rightarrow H_2O^+\rightarrow H_3O^+}$,
     and we also argue that the H$^+$ production is most likely dominated by
     X-ray/cosmic ray ionization.  
     The full set of observations and models leads us to propose that the
     molecular ions arise in a relatively low density ($\gtrsim10^4$ cm$^{-3}$)
     interclump medium, in which case the ionization rate per H nucleus
       (including secondary ionizations) is $\zeta>10^{-13}$ s$^{-1}$, a
     lower limit that is $\mathrm{several}\times10^2$ times the highest
     rate estimates for Galactic regions. In Arp~220, our lower limit for
     $\zeta$ is compatible with estimates for the cosmic ray energy
     density inferred previously from the supernova rate and synchrotron radio
     emission, and also with the expected ionization rate produced by
     X-rays. In NGC~4418, we argue that X-ray ionization due to an AGN is
     responsible for the molecular ion production.   
}


   \maketitle
%

\section{Introduction}

Observations of O-bearing molecular ions are powerful probes of
the oxygen chemistry and related physical processes in the interstellar
medium. The formation of \ohp, \hdop, and \htop\ can be initiated by
cosmic ray and/or X-ray ionization of H and H$_2$ 
\citep[e.g.][]{her73,mal96}, and
hence their abundances are sensitive to the cosmic ray and X-ray fluxes that
permeate the observed environments. Significant amounts of molecular
ions can also be formed in the surfaces of photodissociation regions
  (PDRs) \citep{ste95}. The \ohp/\hdop\ and \ohp/\htop\ 
abundance ratios are sensitive to the molecular fraction and to the fractional
ionization, as the destruction paths for the three species are dominated
  by dissociative recombination (\htop) and reactions with H$_2$ \citep[\ohp\
  and \hdop;][]{ger10,neu10}. In environments with $T<250$ K, these are the
intermediate species for the main gas-phase production route of OH and \hdo,
which are abundant species in the nuclei of some (ultra)luminous infrared
galaxies ((U)LIRGs).  

The successful launch and operation of the Herschel Space Observatory and
  its instruments have 
allowed the observation of O-bearing molecular ions 
in both galactic and extragalactic sources. 
In the Milky Way, and following the first detection of \ohp\ towards
  Sgr~B2 with APEX \citep{wyr10}, 
the ground-state lines of \ohp\ and \hdop\ have been observed in
absorption toward a number of star-forming regions with high submillimeter
continua, mainly tracing intervining diffuse/translucent clouds with low H$_2$
fraction \citep{oss10,ger10,neu10}. The ground-state lines of both species
have also been detected in the Orion BN/KL outflow, though still in absorption
against the continuum \citep{gup10}, and toward the high-mass star forming
region AFGL 2591, also indicating low excitation for both hydrides
\citep{bru10b}. 
Most of the Galactic \ohp\ and \hdop\ detections essentially sample
low-excitation gas and are thought to be generated by the cosmic ray
field in our galaxy \citep{oss10,ger10,neu10,hol12}.
The more stable \htop\ was first detected in the Sgr B2 molecular cloud
  complex near the Galactic center 
\citep{woo91}, and further mapped with APEX in the same region \citep{vdt06}
through the non-metastable $3_2^+-2_2^-$ submillimeter line at 365 GHz. 
Recently, high-lying lines of \htop\ from metastable levels have been detected
in Sgr~B2 by \cite{lis12}, who suggested that the excitation could be due to
formation pumping in X-ray irradiated gas.

The situation in luminous extragalactic sources is different in some
aspects. While the ground-state transition of ortho-\hdop\ has been
detected in absorption in M82 \citep{wei10}, the ground-state \ohp\ and
\hdop\ lines have been detected in emission in Mrk 231, probably indicating a
combination of high excitation conditions and high ionization rates
that are plausibly tracing X-ray Dominated Regions (XDRs) due
to the AGN \citep{vdw10} or the high cosmic ray intensities suggested
  to be present in ULIRGs \citep{pap10}. 
From ISO observations, hints of \ohp\ absorption
at $\sim153$ \microns\ from the first excited rotational level were previously
reported in the ULIRGs Arp 220 \citep[][hereafter G-A04]{gon04}, and Mrk
231 \citep{gon08}, but the absorption in Arp 220 was mostly ascribed to NH. 
Recently, \cite{ran11} (hereafter R11) have reported the detection of the
ground-state \ohp\ and o-\hdop\ fine-structure components in absorption
toward Arp 220, as well as two emission lines of \hdop\ with excited lower
levels. On the other hand, the \htop\ $3_2^+-2_2^-$ submillimeter line has
also been detected in Arp 220 \citep[][hereafter 
vdT08]{vdt08} and other nearby galaxies \citep{aal11}.

We report in this work on the Herschel/PACS detection of excited \ohp,
\hdop, and \htop\ in both Arp 220 and NGC 4418. In a recent study
\citep[][hereafter Paper I]{gon12}, we have reported on the Herschel/PACS
observations of these two sources in \hdo, H$_2^{18}$O, OH, $^{18}$OH, HCN,
and NH$_3$, together with a quantitative analysis in which  
a deconvolution of the continuum into several components enabled us to derive
the column densities and abundances of these species associated with each of
the components. A chemical dichotomy was found in Paper I: on the one hand,
the nuclear regions of both galaxies have high H$_2$O abundances
($\sim10^{-5}$), and high HCN and NH$_3$ column densities, indicating a
chemistry characterized by high \tgas, \tdust, and high density, where
the release of molecules from grain mantles to the gas phase has occurred
(through evaporation in ``hot core'' chemistry and sputtering in shocks).
On the other hand, the high OH abundance found in both
sources, with $\mathrm{OH/H_2O}\sim0.5$, appears to indicate the occurrence of
photoprocesses due to X-rays and/or the effects of cosmic rays. This second
aspect of the chemistry is further studied in this paper from the observed
O-bearing molecular cations detected in the Herschel/PACS spectra of both
sources. The paper is organized as follows: the observations are presented in
Sect.~\ref{sec:obs}; column densities and abundances are inferred in
Sect.~\ref{sec:col}; chemical models are presented and compared with
observations in Sect.~\ref{sec:chem}; the inferred important processes
  are discussed in Sect.~\ref{sec:discussion}, and the main conclusions are
listed in Sect.~\ref{sec:conclusions}. As in Paper~I, we adopt distances
to Arp~220 and NGC~4418 of 72 and 29 Mpc, and redshifts of $z=0.0181$ and
  $0.00705$, respectively.


   \begin{figure*}
   \centering
   \includegraphics[width=14.0cm]{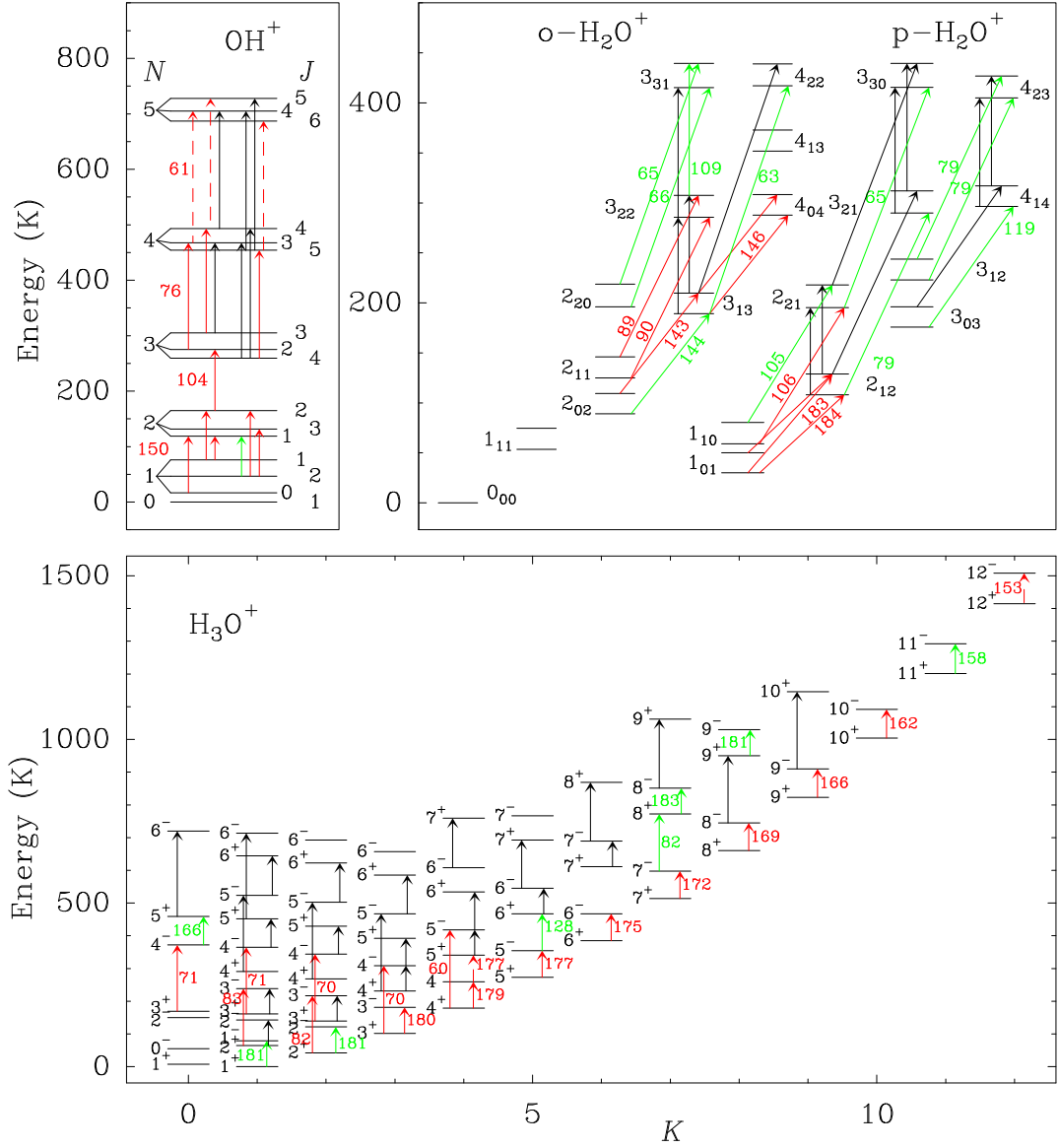}
   \caption{Energy level diagrams of OH$^+$, H$_2$O$^+$, and H$_3$O$^+$,
     showing with arrows the transitions that lie within the PACS range;
     colored numbers indicate round-off wavelengths in $\mu$m. The
     fine-structure splitting of the OH$^+$ and H$_2$O$^+$ levels is out of
     scale. Solid-red arrows mark the lines detected in NGC~4418 and/or
     Arp~220; dashed-red arrows indicate marginal detection in any of the
     sources; green arrows indicate blended lines, and black arrows show
     undetected lines.}
    \label{ener}
    \end{figure*}

\section{Observations and results}
\label{sec:obs}

The full, high resolution, PACS spectra of NGC 4418 and Arp 220, were
  observed as part of the guaranteed-time key program SHINING on July
27th and February 27th (2010), respectively\footnote{The observation
    identification numbers (OBSIDs) are $1342202107-1342202116$ for NGC~4418,
    and $134191304-1342191313$ and $1342238928-1342238937$ for Arp~220.}.  
The data reduction was done using
the standard PACS reduction and calibration pipeline (ipipe) included in HIPE
5.0 975. For the final calibration, the spectrum was normalized to the
telescope flux and recalibrated with a reference telescope spectrum
obtained from Neptune observations. Further details of the data
reduction are given in \cite{gra11}, \cite{stu11}, and Paper I. 

The PACS spectrum of Arp~220 in Paper I was corrected in both flux and
wavelength calibration due to the effects of sub-spaxel pointing errors. 
Following newly made software corrections to the Herschel pointing star
catalogue and the star-tracker x and y focal lengths, a second full spectrum
of Arp 220 was taken by the Observatory to assess the improvement to
Herschel's pointing.  Although the baseline of the new spectrum is of higher
quality, general agreement with our corrected original spectrum was found in
both flux and wavelength calibration. The data shown here are mostly an
average of the two observations, except for some ranges where the new data set
yields a more reliable baseline.

Spectroscopic parameters of \ohp, \hdop, and \htop\ used for line
identification and radiative transfer modeling were mostly taken from the
spectral line catalogs of the CDMS \citep{mul01,mul05} and JPL \citep{pic98},
and are included in Table~\ref{tab:fluxes}, together with the equivalent
  widths ($W_{\mathrm{eq}}$) and fluxes of the detected lines. The wavelength
  uncertainties tabulated in the catalogs are significantly lower than the
  PACS spectral resolution.  Energy
level diagrams for the three cations are shown in Fig.~\ref{ener}, where the
solid red arrows indicate the detected transitions in at least one of the
sources.

\subsection{\ohp}
\label{obs:oh+}

Owing to the electronic spin $S=1$, all rotational
levels of \ohp\ except the ground $N=0$ state are split into 3
fine-structure levels with total angular momentum $J=N-1,N,N+1$
(Fig.~\ref{ener})\footnote{The hyperfine structure of both \ohp\ and \hdop\ is
  ignored in this work as the line splitting is 
much lower than both the PACS spectral resolution and the linewidths.}. 
Of the six allowed fine-structure transitions between
excited rotational levels, the three with $\Delta J=\Delta N$ are the
strongest. Line spectroscopic parameters of \ohp\ were taken
from the CDMS, and were derived by \cite{mul05}. In that work, the $N =
1 - 0$ transition frequencies from \cite{bek85} were fit together with
additional infrared data. A dipole moment of 2.256~D was taken from an
\textit{ab initio} study \citep{wer83}.

\begin{figure*}
   \centering
   \includegraphics[width=17.0cm]{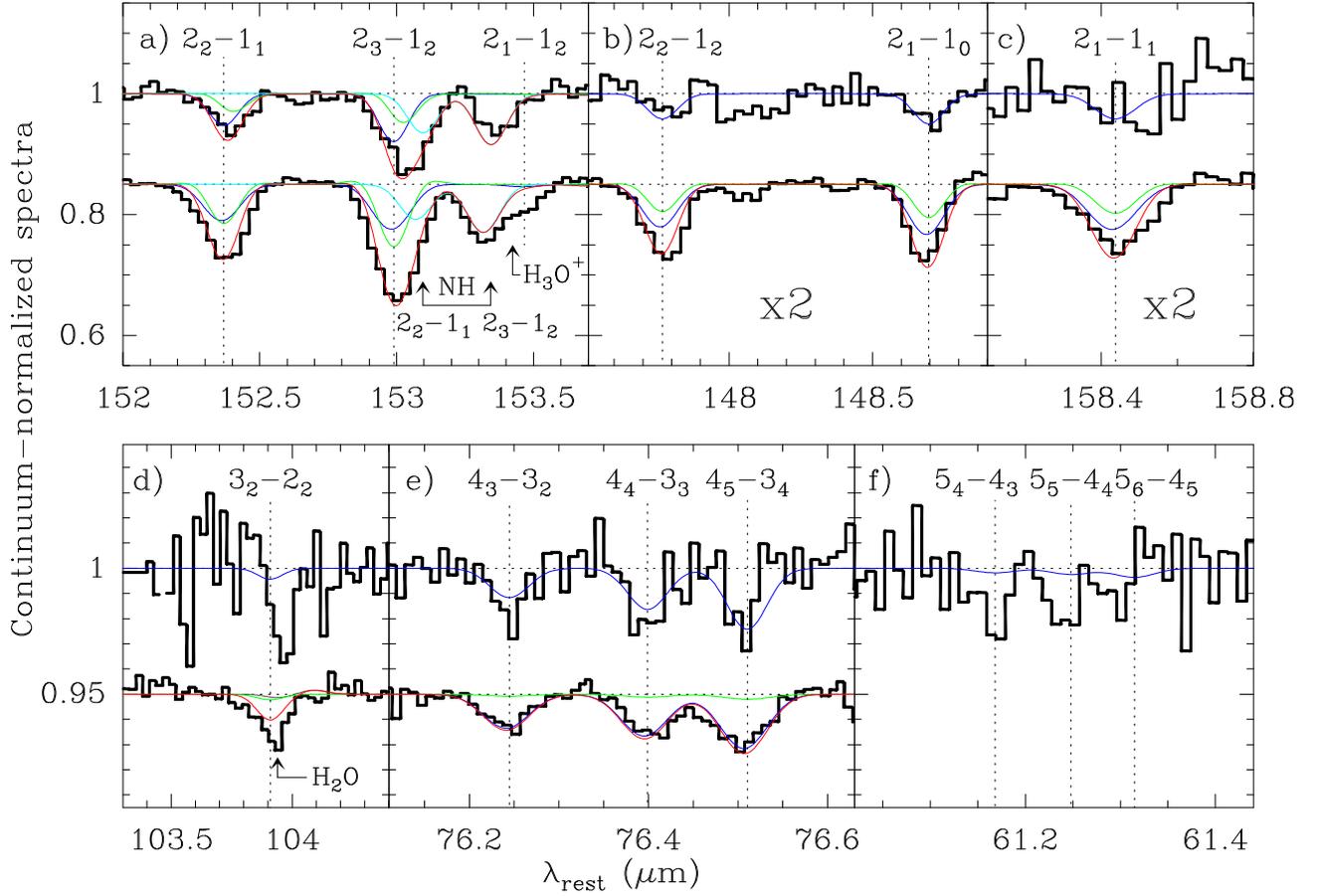}
   \caption{Continuum-normalized spectra around the OH$^+$ lines in NGC~4418
     (upper histograms) and Arp~220 (lower histograms). The dotted
       vertical lines indicate the rest wavelengths of the transitions for the
       nominal redshifts of $z=0.00705$ (NGC~4418) and $0.0181$ (Arp~220).
     The spectra in 
     panels {\bf b} and {\bf c} have been scaled by a factor 2. (Potential)
     contamination by other species (NH in panel {\bf a} and H$_2$O in {\bf d})
     is also indicated. The marginal $5_{J}-4_{J'}$ lines in {\bf f}
     correspond to NGC~4418; this part of the spectrum is not available in
     Arp~220. Blue lines show model fits for \ohp\ in the nuclear
     components of both NGC~4418 and Arp~220, the green lines show the
     model fit for \ohp\ in the extended components of both sources, and
     the light-blue lines in {\bf a} are model fits for NH; red is total. } 
   \label{spectraoh+}
    \end{figure*}

Figure~\ref{spectraoh+} shows the observed spectra around the  
\ohp\ lines in NGC 4418 (upper histograms) and Arp 220 (lower histograms). 
The spectra around the positions of the $2_{J}\leftarrow 1_{J'}$ lines, with
$E_{\mathrm{lower}}\approx50$ K, are displayed in panels a-c. 
The strongest \tohplus2312\ at $\approx153$ \microns\  
is partially blended with NH, but the \tohplus2211\ is free from contamination
and clearly detected in both sources. In Arp 220, the 153 \microns\ absorption
appears to be dominated by \ohp\ rather than by NH, in contrast with the
initial estimate by \cite{gon04} based on ISO data. The \tohplus2110\ at
$\approx148.7$ \microns\ is also well detected, and the intrinsically weaker
\tohplus2111\ ($158.4$ $\mu$m) and \tohplus2212\ ($147.8$ $\mu$m) lines are
(marginally) detected in NGC 4418 but still very strong in Arp 220. There is
in addition an apparent red-shifted wing in the NH \tohplus2312\
spectrum of Arp 220, detected in both data sets, which coincides with the
position of the intrinsically weak \ohp\ \tohplus2112\ transition. As shown
below (Sect.~\ref{obs:h3o+}), however, this wing feature is most probably due
to the very excited $12_{12}^{-}\leftarrow12_{12}^{+}$ line of \htop. 

Most $3_{J}\leftarrow 2_{J'}$ lines at $\sim100$ \microns\ lie within the gap
between the green ($\lambda<100$ $\mu$m) and red ($\lambda>100$ $\mu$m) bands
of PACS and are not observable, except for the \tohplus3222\ transition at
$103.9$ $\mu$m shown in Fig.~\ref{spectraoh+}d. The line in NGC 4418 is close
to the edge of the red band where the spectral noise is high, but in Arp 220 a
clear spectral feature is detected at the expected wavelength. The line,
however, is blended with the \hdo\ \t615606\ transition (Paper I), which may
account for some of the observed absorption.

The three strongest $4_{J}\leftarrow3_{J'}$ lines at $76-76.5$ \microns\
are clearly detected in both sources (Fig.~\ref{spectraoh+}e), indicating a
high-excitation component in the transient \ohp. It is worth noting that while
the $2_{J}\leftarrow1_{J'}$ lines are stronger in Arp 220, the more excited
$4_{J}\leftarrow3_{J'}$ lines are slightly deeper in NGC 4418, resembling the
behavior found for other species like \hdo, OH, HCN, and NH$_3$ (Paper I). The
$4_{J}\leftarrow3_{J'}$ lines still have slightly higher equivalent widths in
Arp~220 (Table~\ref{tab:fluxes}) owing to the broader line profiles in this
source. The line opacities of the fine-structure components are 
$\tau\propto \lambda^3g_u A_{ul}$ if the lower levels are populated according
to their degeneracies, and then
$\tau_{4_5-3_4}:\tau_{4_4-3_3}:\tau_{4_3-3_2}=1:0.77:0.58$. These are expected
to be the ratios of the line equivalent widths ($W_{\mathrm{eq}}$) if the three
components are optically thin. In Arp~220, the observed $W_{\mathrm{eq}}$ ratios
are $1:(0.69\pm0.11):(0.53\pm0.10)$ (Table~\ref{tab:fluxes}),
consistent with optically thin absorption. In
NGC~4418 they are $1:(1.2\pm0.6):(1.0\pm0.5)$, compatible within the
  uncertainties with both optically thin and saturated absorption.

In NGC 4418, some marginal features around 61 \microns\ roughly
coincide with the positions of the \ohp\ $5_{J}\leftarrow4_{J'}$ lines
(Fig.~\ref{spectraoh+}f). However, the feature associated with the
intrinsically strongest \tohplus5645\ component is relatively weak and shifted
in wavelength by $0.015$ \microns, shedding doubt on this
tantalizing identification. In Arp~220, this part of the spectrum was affected
by instrumental problems and is not shown.

\subsection{H$_2$O$^+$}
\label{obs:h2o+}

   \begin{figure*}
   \centering
   \includegraphics[width=17.0cm]{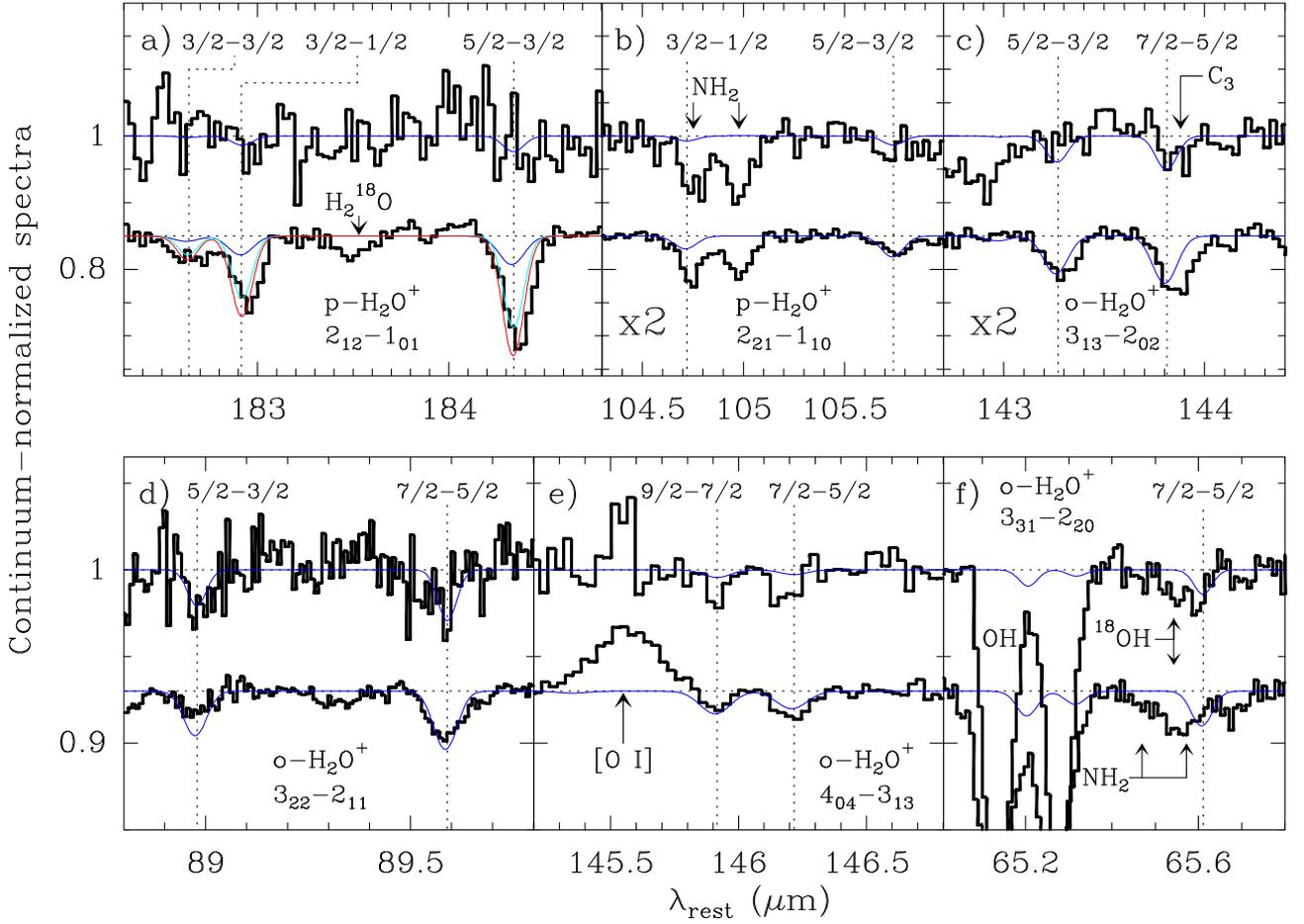}
   \caption{Continuum-normalized spectra around the H$_2$O$^+$ lines in NGC
     4418 (upper histograms) and Arp 220 (lower histograms). The dotted
       vertical lines indicate the rest wavelengths of the transitions for the
       nominal redshifts of $z=0.00705$ (NGC~4418) and $0.0181$ (Arp~220). 
     The spectra in 
     panels {\bf b} and {\bf c} have been scaled by a factor 2. Close features
     due to other species and potential contaminations are indicated. Dark
     blue and light blue lines show model fits; red is total.}  
    \label{spectrah2o+}
    \end{figure*}

   \begin{figure*}
   \centering
   \includegraphics[width=17.0cm]{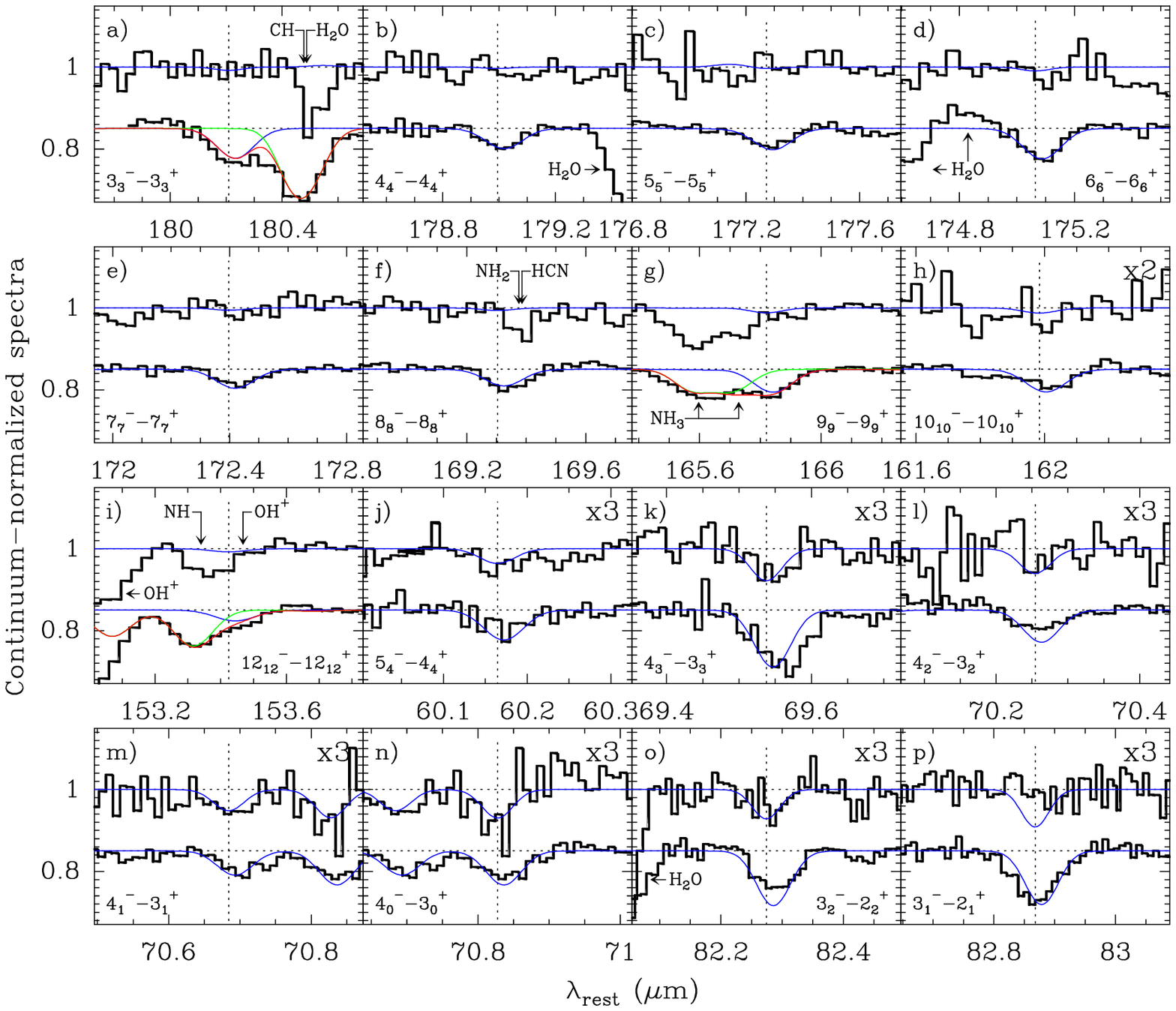}
\caption{Continuum-normalized spectra around the \htop\ lines in NGC~4418
  (upper histograms) and Arp~220 (lower histograms). The dotted vertical
    lines indicate the rest wavelengths of the transitions for the nominal
    redshifts of $z=0.00705$ (NGC~4418) and $0.0181$ (Arp~220).
  Panels {\bf a}-{\bf i} show the pure inversion, metastable lines of \htop,
  while panels {\bf j}-{\bf p} show the rotation-inversion lines at
  $\lambda<100$ $\mu$m. Numbers below line labeling indicate wavelengths in
  $\mu$m. The spectra in panels {\bf h} and {\bf j}-{\bf p} have
  been scaled by a factor 2 and 3, respectively. 
  The $1_1^--1_1^+$ and $2_2^--2_2^+$ lines at 
  $\approx181$ $\mu$m are strongly contaminated by lines of CH and
  H$_2^{18}$O, and are not shown. Close features due to other species and
  potential contaminations are indicated. In panels {\bf f} and {\bf i}, the
  contamination by NH$_2$, HCN, and OH$^+$ is expected to be small (see
  text). The blue lines show the best-fit model for \htop; the green
  lines in panels {\bf g} and {\bf i} show model fits for NH$_3$ and NH lines;
  red is the total.}  
\label{spectrah3o+}
\end{figure*}

The asymmetric \hdop\ also has fine-structure splitting of the rotational
levels producing doublets ($S=1/2$) and yielding a very rich
spectrum in the far-infrared domain (Fig.~\ref{ener}). 
Predictions of the rotational spectrum were taken from the CDMS.
The CDMS data were based on extrapolated zero-field frequencies in \cite{mur98},
the derived $N = 1 - 0$ transition frequencies based on \cite{str86}, and
additional rovibrational (infrared) transitions and ground state 
data from electronic spectra. 
HIFI observations of the $N =  1_{10} - 1_{01}$ transition frequencies above
600~GHz toward Sgr~B2(M) reported by \cite{sch10} are compatible with the
transition frequencies in the CDMS, although these lines lie outside our
accessible frequency range.
A ground state dipole moment of $2.398$ D was calculated \textit{ab initio} by
\cite{wei89}.

As shown in Fig.~\ref{ener}, most low-lying lines of \hdop\ are
blended with lines of other species, but some of them lie in free
windows (Fig.~\ref{spectrah2o+}). Very prominent is the ground-state
p-\hdop\ \t212101\ transition at $183-184$ \microns\ in Arp 220, with the three
fine-structure components clearly identified in the spectrum. This
transition is not detected in NGC 4418. In Arp~220,
the observed line strengths strongly decrease with increasing
$E_{\mathrm{lower}}$, indicating that the ground-state p-\hdop\ lines arise
in low excitation gas. The opacity ratios of these components are 
$\tau_{5/2-3/2}:\tau_{3/2-1/2}:\tau_{3/2-3/2}=1:0.55:0.12$, so the observed
$W_{\mathrm{eq}}$ ratios of $1:(0.65\pm0.07):(0.30\pm0.05)$
(Table~\ref{tab:fluxes}) suggest line saturation effects in at least the
strongest $5/2-3/2$ component. For $\sigma_v=60$ \kms, this component attains
an optical depth of unity with a column of
$N(\mathrm{p-H_2O^+})\approx5\times10^{14}$ \cmd. The line absorbs only
$\approx15\%$ of the continuum, likely indicating 
some spectral dilution and a low covering factor of the continuum.

The two prominent
features at $104.7-105$ \microns\ in both NGC~4418 and Arp~220 are most
probably due to NH$_2$. The feature at $143.8$ \microns\ may be 
contaminated by C$_3$. In Arp~220, the excited \t221110, \t313202,
\t322211, and \t404313\ lines of \hdop, with $E_{\mathrm{lower}}$ up to $190$
K, are clearly detected, but in NGC~4418 only the high-lying \t322211\
transition show measurable absorption.
Detection of the \t404313\ transition in NGC~4418 is questionable, 
as the spectral features associated with the $9/2-7/2$ and $7/2-5/2$
components are either too narrow or shifted (Fig.~\ref{spectrah2o+}e).

\subsection{H$_3$O$^+$}
\label{obs:h3o+}

H$_3$O$^+$ and \nht\ are isoelectronic and spectroscopically similar: both are
oblate symmetric top rotors, have a double-minimum potential such that inversion
takes place between the two equivalent minima, and have \textit{ortho}
($K=3n$) and \textit{para} ($K\ne 3n$) modifications. 
Fig.~\ref{ener} shows the energy level diagram for \htop.
Owing to the $\Delta K = 0$ selection rule for all radiative transitions, the
$J = K$ levels in the lower tunneling state 
cannot be radiatively pumped and hence are metastable. They
therefore will be "thermally" populated at a rotational temperature, \trot,
which reflects either \tgas\ or, if the formation-destruction rates are faster
than the rate for collisional relaxation, the formation process characteristics
\citep[e.g.][see also Appendix~\ref{appb}]{bru10a,lis12}.

Spectroscopic data for \htop\ were taken from the JPL catalogue
\citep{pic98}. The entry was based on the recent analysis by \cite{yu09}, who
also presented some improved transition frequencies. The majority of the pure
tunneling transition frequencies were reported by \cite{ver89}. 
A ground state dipole moment of $1.44$ D was calculated \textit{ab initio} by
\cite{bot85}. The antisymmetric\footnote{The symmetric and antisymmetric 
  states are here denoted with $+$ and $-$, respectively.} state is higher in
energy by 1659~GHz \citep{yu09} because of the low barrier to linearity,
and thus the pure inversion transitions of \htop\ lie in the far-IR domain
($150-185$ $\mu$m)\footnote{For \nht, the antisymmetric state is higher in
  energy by 23.7~GHz and the pure-inversion transitions lie at centimeter
  wavelengths.}.

The solid-red arrows in Fig.~\ref{ener} indicate the \htop\ lines detected in
Arp~220, and Fig.~\ref{spectrah3o+} presents the observed \htop\ spectra
  of NGC~4418 and Arp~220. In Arp~220,
spectral features at the wavelengths of the \htop\ pure-inversion, metastable 
(hereafter PIMS) lines (panels a-i) are detected up to a lower level energy of
$\approx1400$ K above the ground state ($12_{12}^{+}$), while none of these
lines is detected in NGC~4418. 
The rotation-inversion (hereafter RI) lines lie at shorter wavelengths
  ($\lambda<100$ $\mu$m) and are shown in Fig.~\ref{spectrah3o+}j-p. Four of
  them arise from non-metastable levels (Fig.~\ref{ener}), three of them from
  metastable levels (hereafter RIMS lines), and are detected within the lowest
  $K\leq4$ ladders up to an energy of $\sim200$ K in both galaxies.

The $1_{1}^{-}-1_{1}^{+}$, $2_{2}^{-}-2_{2}^{+}$, and 
$11_{11}^{-}-11_{11}^{+}$ PIMS lines are contaminated by strong lines of
CH, H$_2^{18}$O, and [C {\sc ii}], and are not shown. All features in
Arp~220 are detected in both datasets. Close lines due to other species,
as well as potential contaminations (panels f and i), are also indicated in
Fig.~\ref{spectrah3o+}. Based on the weakness of the HCN $19-18$ line (Paper I),
the HCN $20-19$ line at $169.39$ $\mu$m is not expected to strongly
contaminate the \htop\ $8_{8}^{-}\leftarrow8_{8}^{+}$ line, though some
contribution from HCN is not ruled out (see below). Likewise, the nearby NH$_2$
\t330321\ $4-3$ line is intrinsically weak, and NH$_2$ modeling that accounts
for other detected NH$_2$ lines does not produce significant absorption at
$169.38$ $\mu$m. A broad feature is detected at
$165.7$ $\mu$m (panel g), which cannot be explained by only the NH$_3$
$3_{2}^{-}\leftarrow2_{2}^{+}$ and $3_{1}^{-}\leftarrow2_{1}^{+}$ lines (Paper
I); thus strong absorption by ortho-\htop\ $9_{9}^{-}\leftarrow9_{9}^{+}$ is
inferred. The relatively weak \htop\ $10_{10}^{-}\leftarrow10_{10}^{+}$ line
is affected by an uncertain baseline, 
though the line is detected in both data sets. On the other hand, a line wing
feature close to the NH $2_3\leftarrow1_2$ $153.35$ $\mu$m line is detected,
coinciding with both the ortho-\htop\ $12_{12}^{-}\leftarrow12_{12}^{+}$ and
the OH$^+$ $2_1\leftarrow1_2$ lines ($153.47$ $\mu$m, see also
Fig.~\ref{spectraoh+}a). As mentioned earlier, the latter is also 
intrinsically weak, and modeling of the other OH$^+$ $2_{J}\leftarrow1_{J'}$
lines (Sect.~\ref{sec:arp220models}) yields negligible absorption in that
component. Upon inspection of the CDMS and JPL catalogs, we have not found any
other reliable candidate for the NH wing, which we tentatively ascribe to the
very high-lying \htop\ line.

No potential contamination has been found for the RI lines. Contrary to the
PIMS lines, some of these RI lines are detected in NGC~4418. The
non-metastable levels are populated through absorption of far-IR photons,
and trace the local far-IR radiation density.

\begin{figure}[h]
   \centering
   \includegraphics[width=8.5cm]{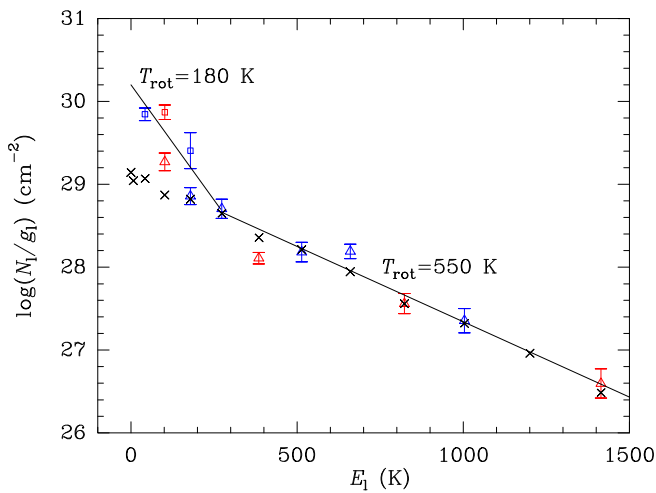}
\caption{Population diagram of \htop\ (metastable levels) in Arp~220. Red and
  blue symbols correspond to ortho and para levels, respectively. Columns
  derived from pure-inversion and rotation-inversion lines are shown with
  triangles and squares, respectively. The crosses indicate the expected
  population distribution for $T_{\mathrm{rot}}=550$ K and a fixed
  ortho-to-para ratio of 1.}   
\label{diagh3o+}
\end{figure}

The PIMS and RIMS lines are observed in absorption
against the strong far-IR continuum source of Arp~220, and their relative
strengths probe \trot. The population diagram of \htop\ is shown in
Fig.~\ref{diagh3o+}, where the level columns have been calculated in the
optically thin limit and by assuming that the \htop\ covers the whole
continuum from Arp 220. Since the degeneracy of the ortho levels is twice that
of the para levels, we expect a nearly continuous distribution for an
ortho-to-para ratio of 1, as observed in the high-lying levels (crosses in
Fig.~\ref{diagh3o+}). The distribution is roughly described with two values of
\trot, $\sim180$ K up to the $5_5^+$ level, and $\sim550$ K for higher-lying
levels. This is strikingly similar to that found by \cite{lis12} towards
Sgr~B2. The column of the $8_8^+$ level ($E_l\approx660$ K) appears to be
higher than expected, suggesting some contamination by HCN at $169.4$
$\mu$m (Fig.~\ref{spectrah3o+}f).  
On the other hand, there are two metastable levels, the $3_3^+$ and the
$4_4^+$, for which both PIMS and RIMS lines are available. In both levels, the
columns derived from the RIMS lines are higher by a factor of $\approx1.8$
than those derived from the PIMS lines. Two main reasons can account for this:
first, the level columns are calculated, as mentioned above, by assuming that
the \htop\ covers the whole far-IR continuum, which is relatively cold
(i.e. $S_{180}/S_{60}=0.43$). If the \htop\ is absorbing a warmer
  continuum that is diluted within the observed SED (i.e. filled in by a
  cooler continuum source), the intrinsic equivalent width and
column of the PIMS lines will be increased relative to those of the RIMS line,
and both columns will be in better agreement. Therefore, the discrepancy
probably indicates that the highly excited \htop\ covers only a fraction of the
observed continuum, and that the level columns are lower limits.  
Second, the columns in Fig.~\ref{diagh3o+} are not corrected
for stimulated emission, which will mainly increase the columns of the
PIMS lines at long wavelengths.

\subsection{Velocity shifts in Arp 220}
\label{sec:vel_arp220}

\begin{figure*}
   \centering
   \includegraphics[width=15cm]{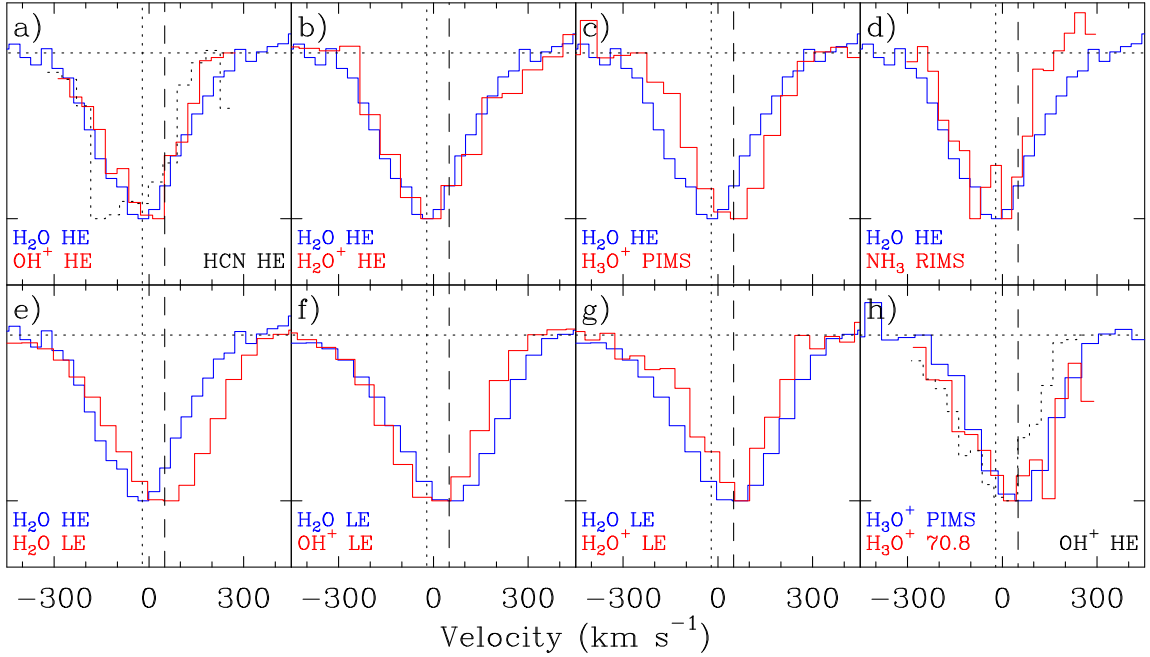}
\caption{Comparison between averaged line shapes of several lines in Arp 220. 
  HE, LE, PIMS and RIMS mean high-excitation, low-excitation, pure-inversion
  metastable, and rotation-inversion metastable, respectively. 
  \hdo\ HE is the average of the \t616505\ and \t322211\ lines of \hdo (Paper
  I);  
  HCN HE is the $18-17$ line of HCN;
  \ohp\ HE is the average of the $4_{3}-3_{2}$ and $4_{5}-3_{4}$ lines of \ohp;
  \hdop\ HE is the average of the \t322211\ $7/2-5/2$, \t313202\ $5/2-3/2$,
  and \t404313\ $7/2-5/2$ lines of \hdop; 
  \htop\ PIMS is the average of the pure-inversion, metastable 
  $5_{5}^{-}-5_{5}^{+}$, $6_{6}^{-}-6_{6}^{+}$, $7_{7}^{-}-7_{7}^{+}$, and
  $8_{8}^{-}-8_{8}^{+}$ lines of \htop;
  \nht\ RIMS is the average of the rotation-inversion metastable
  $7_{6}^{+}-6_{6}^{-}$ and $6_{5}^{-}-5_{5}^{+}$ lines of \nht;
  \hdo\ LE is the average of the \t212101\ and \t221110\ lines of \hdo;
  \ohp\ LE is the average of the $2_{2}-1_{2}$, $2_{1}-1_{0}$, $2_{2}-1_{1}$,
  and $2_{1}-1_{1}$ lines of \ohp; 
  \hdop\ LE is the average of the \t212101\ $5/2-3/2$ and $3/2-1/2$ lines of
  \hdop;
  \htop\ 70.8 is the $4_{0}^{-}-3_{0}^{+}$ line of \htop\ at 70.8 \microns.
The dotted and dashed vertical lines indicate the velocity centroids of
  the averaged HE and LE spectra, respectively.
}  
\label{vel}
\end{figure*}

\begin{figure}
   \centering
   \includegraphics[width=8.5cm]{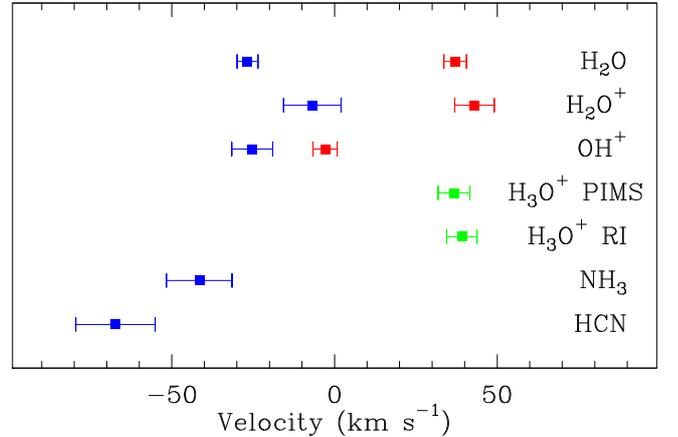}
\caption{Velocity centroids of the averaged line shapes in Arp~220
  displayed in Fig.~\ref{vel}, derived from Gaussian fits. 
  Velocities are calculated relative to $z=0.0181$.
  Blue and red symbols correspond 
  to HE and LE averages, respectively; for \htop\ the green symbols
  indicate the PIMS and RI averages. Errorbars are 1$\sigma$ uncertainties.
}  
\label{velcenter}
\end{figure}

Having confirmed the velocity calibration (or "corrections") applied to
  the Arp~220 spectrum (Paper I) with the second Arp 220 dataset, we now
  compare the line profiles and velocity shifts presented in Paper~I to the
  velocity profiles and centroids of the molecular ions.
The average profiles of several \hdo, \ohp, \hdop, \htop, HCN and
\nht\ lines detected in Arp 220 are compared in Fig.~\ref{vel}. In all panels,
HE and LE mean high-excitation and low-excitation, respectively,
qualitatively indicating whether the selected lines are high-lying or
low-lying lines. Thus \hdo\ HE is the average profile of two representative
high-excitation lines of \hdo, the \t616505\ and \t322211\ lines (Paper I); 
\hdo\ LE is the average profile of the two low-lying \t212101\ and \t221110\
lines, and similarly for other species/excitation. 
The \htop\ PIMS spectrum is the average of the unblended high-lying,
pure-inversion metastable lines of \htop, and the \nht\ RIMS one is the
average of the relatively uncontaminated rotation-inversion metastable
$7_{6}^{+}\leftarrow6_{6}^{-}$ and $6_{5}^{-}\leftarrow5_{5}^{+}$ lines. The
averages were generating by resampling the individual spectra to a common
velocity resolution, and adopting equal weights for all them. 
Figure~\ref{velcenter} displays the velocity centroids of these generated line
profiles. 

Figures~\ref{vel} and \ref{velcenter} indicate that there are systematic
velocity shifts associated with line excitation, irrespective of the
considered species. The averaged \hdo\ HE profile, peaking at
$-20$ \kms, has a shape very similar to the averaged \ohp\ HE (panel a),
\hdop\ HE (b), and \nht\ 
HE (d) profiles, though the latter is somewhat narrower (and relatively
uncertain due to line blending). The HCN $18\leftarrow17$ line (panel a) could
be even more blueshifted, though 
the signal-to-noise ratio is relatively low. On the other hand,
there is a clear velocity shift in the peak absorption of the \hdo\ HE and LE
averages, with the latter peaking at a velocity of $+50$ \kms. The LE averages
of \ohp\ and \hdop\ also tend to peak at more positive velocities, though the
\ohp\ LE lines are not ground-state lines ($E_{\mathrm{lower}}\approx50$ K)
and thus may represent an intermediate situation.

The velocity centroids of the \htop\ PIMS lines, however, which are
  characterized by high $E_{\mathrm{lower}}$, peak at the velocity of the LE
  lines and thus present an exception to the finding discussed above.
A velocity shift of $\approx70$ \kms\ is shown in Fig.~\ref{vel}c between the
average \hdo\ HE and \htop\ PIMS spectra, with the latter peaking at the
velocity of the LE lines of \hdo\ and \hdop. 
This shift is much lower than the line widths of $\sim300$ \kms, but 
larger than our spectral calibration accuracy.
The average velocity of all \hdo\ lines (excluding the
lowest-lying LE lines) is $-20\pm35$ \kms, whereas that of the metastable
\htop\ lines is $50\pm25$ \kms. The latter corresponds to
$v_{\mathrm{LSR}}=5480$ \kms, in rough agreement with the velocity centroid of
the $3_{2}^{+}-2_{2}^{-}$ line (vdT08) and close to the central velocity of CO
2-1 in the western nucleus \citep{dow07}.
Excluding the low signal-to-noise \htop\ $4_{1}^{-}\leftarrow3_{1}^{+}$ line, 
the other three non-metastable lines have an average velocity of $20\pm15$
\kms. The low-lying $4_{3}^{-}\leftarrow3_{3}^{+}$ line
(Fig.~\ref{spectrah3o+}k) is even more redshifted ($\approx100$ \kms), with
the two data sets giving very similar line shapes. With the same lower level,
the pure-inversion $3_{3}^{-}\leftarrow3_{3}^{+}$ line
(Fig.~\ref{spectrah3o+}a) also shows hints of redshifted absorption at $>100$
\kms.

At the very least, the velocity shift between the peak absorption of the
\htop\ PIMS and the HE lines of other species indicates a {\em spatial} shift
between the regions responsible for the strongest absorption in both sets of
lines. On the one hand, this may not be surprising from excitation
arguments. The \hdo\ HE levels are excited through radiative pumping by dust
(Paper I), and thus the corresponding lines trace the region of highest
far-IR radiation density (and thus of highest \tdust). On the other hand, the
\htop\ metastable levels cannot be radiatively pumped, but are either
excited through collisions indicating high \tgas, or reflect
\htop\ formation in high-lying levels; thus the physical conditions
that favor strong absorption in each set of lines are different and 
  do not necessarily coincide spatially. On the other hand, the 
\nht\ RIMS lines, with $E_{\mathrm{lower}}=300-400$ K, are collisionally
pumped but better match the \hdo\ HE profile (Fig.~\ref{vel}d), 
suggesting either chemical differences between the two velocity
components, or the dominance of chemical pumping on the formation of the PIMS
lines of \htop.

Given the velocity coincidence between the peak absorption of the \htop\ PIMS
lines and the \hdo\ LE lines, the question is whether the \htop\ PIMS lines
arise from foreground gas in the direction of, but detached from the nuclear
region, or the lines arise on the contrary from gas physically associated with
the nuclei. The PACS spectrum of Arp~220 shows RI lines of
\htop\ arising from non-metastable levels (Fig.~\ref{spectrah3o+}m,l,n,p) that 
are radiatively pumped through dust emission. Of the above lines, the
\htop\ $4_{0}^{-}\leftarrow3_{0}^{+}$ at $70.8$ \microns\ is most sensitive to
\tdust, requiring relatively high far-IR radiation densities.
Figure~\ref{vel}h indicates a good match between the shape of the \htop\
$70.8$ \microns\  line and of the PIMS average, and thus it is likely that
both sets of lines arise from essentially the same region. We thus
expect that the \htop\ PIMS lines are of primarily nuclear origin 
though, as argued in Sect.~\ref{sec:mod:h3o+arp220}, probably more extended
than the region where the \nht\ lines are formed.

The simplest scenario that can account for the observed velocity pattern in
Arp~220 consists of the \htop\ PIMS lines and the LE lines of other species
peaking at the ``rest'' velocity of the western nucleus, and thus primarily
tracing rotating gas together with more foreground gas, while the HE lines of
\hdo, OH, \nht, and also \ohp\ and \hdop, mainly tracing an outflowing
component with a line of sight velocity of $\sim75$ \kms, similar to
that observed in HCO$^+$ 
\citep{sak09}\footnote{In Fig. 12 of Paper I, the HCO$^+$ lines should be
  shifted 100 \kms\ to the red, as the original spectrum \citep{sak09} uses
  the radio convention for the velocities; thus the HCO$^+$ absorption toward
  the western nucleus better matches the blueshifted absorption in the \hdo\
  HE lines.}. Besides grain mantle evaporation and neutral-neutral reactions
at high \tgas\ (Paper I), sputtering in shocks could also remove the mantles
and contribute to enhance the gas-phase abundances of \hdo\ 
and NH$_3$. Alternatively, both components may be tracing different  
sections of the nuclear rotating disk(s), which shows steep velocity
gradients as seen in the (sub)millimeter lines of CO \citep{dow07,sak08}.

\section{Environments, column densities and abundances}
\label{sec:col}

\subsection{The excitation of the molecular ions}
\label{sec:excit}

Derivation of the column densities implied by the observations require
calculations of excitation appropiate for the environments where \ohp, \hdop, 
and \htop\ reside. The rates for collisional excitation of these
  molecular ions are not known, but  
both Arp 220 and NGC 4418 have luminous-compact
nuclear far-IR continuum sources that excite \hdo, OH, and other
species to level energies of $>500$ K (Paper I). Therefore, a natural
mechanism for exciting \ohp, \hdop, and the non-metastable levels of \htop\
along a given $K-$ladder is via radiatively allowed transitions through
absorption of dust-emitted photons. 

Collisional excitation of the high-lying lines of \ohp\ and \hdop\  
is expected to be less efficient. The $A_{ul}-$Einstein coefficients are high  
(Table~\ref{tab:fluxes}), and thus the critical densities ($n_{\mathrm{cr}}$)
are likely to be high as well. For the $N=1-0$ transition of \ohp, with
$A_{ul}\approx0.02$ s$^{-1}$ and assuming a coefficient for collisional
de-excitation with H and H$_2$ of $10^{-9}$ cm$^3$ s$^{-1}$, 
$n_{\mathrm{cr}}\approx2\times10^7$ cm$^{-3}$, which is much higher than the
expected densities (see Sect~\ref{sec:zeta} below), and also higher than the
nuclear densities inferred from other tracers (Paper I). As a result, the
expected excitation temperature of the $N=1-0$ transition of \ohp\ from
collisional excitation alone would be $<20$ K, in contrast with the typical
values of $35-70$ K obtained from radiative excitation in the models shown
below. On the other hand, excitation through collisions with electrons has
rate coefficients that might be much higher than with H and H$_2$,
$\sim10^{-6}$ cm$^3$ s$^{-1}$ \citep{neu89,lim99}, and has been recently
invoked by \cite{vdt12} to explain the emission in the HF $1-0$ line toward
the Orion Bar. However, the expected electron density is
$\mathrm{several}\times10^3$ lower than the density of H  
(Fig.~\ref{chemres}), so that its effect on the excitation is not expected to
be much higher than that of H. Finally, the detected lines are seen
in absorption, and radiative pumping models generate the observed absorption
features naturally, as we show below.

Another potential excitation mechanism is chemical formation in
excited levels \citep{sta09}, i.e. formation pumping. In environments with
weak dust emission, a downward cascade after molecular formation in excited
levels \citep[e.g.][]{bru10a} is expected to generate emission in
non-metastable lines (i.e. chemoluminiscence). But again, in the
specific case of NGC~4418 and Arp~220, all detected lines from
molecular ions are seen in absorption, qualitatively matching the far-IR
pumping scheme. Furthermore, ignoring excitation by dust emission, the
equilibrium level populations of \ohp\ and \hdop\ resulting from formation
pumping \citep[see Appendix~B in][]{bru10b} yield very low excitation even for
very high formation rates (see Appendix~\ref{appa}). This is due to the very
high A-Einstein coefficients (Table~\ref{tab:fluxes}) as compared with the
expected formation/destruction rates (Sect.~\ref{sec:chem}). Thus we do not
expect that formation pumping has a measurable effect on the excitation of
non-metastable transitions in these sources.  

The case of the metastable lines of \htop\ is different. The $K>1$ ladders 
cannot be populated from $K=0,1$ through radiative pumping, but only
through collisions or via molecular formation in high-lying levels. The
latter process can be envisioned as follows: molecular formation in an excited
$(K,J)$ level will be followed by quick cascade down to the metastable $J=K$
one, where no more paths for radiative decay are available. Then, the
molecule will ``freeze-out'' in that metastable level (ignoring here
  radiative excitation for simplicity) until destruction through
recombination or until a collisional thermalization event takes place. If the
destruction rate dominates over collisional relaxation, the observed
population distribution in the metastable levels (i.e., \trot) will be
governed by the formation process rather than by \tgas\
\citep{lis12}. Writing the rate of \htop\ formation per unit volume as
$\Gamma_{\mathrm{H_3O^+}}=\epsilon_{\mathrm{H_3O^+}}n_{\mathrm{H}}\zeta$,
where $\epsilon_{\mathrm{H_3O^+}}$ is the efficiency with which ionizations are
transfered to the production of \htop\ \citep[see][for \ohp]{neu10}, the
  condition for formation pumping to dominate the observed excitation over
  collisions can be written as\footnote{In steady state, this condition can
    also be expressed as $k_{\mathrm{col}} < 10^{-10} \times
    (T_{\mathrm{gas}}/100)^{-0.5} \times  
    (\chi(\mathrm{e^-})/1.4\times10^{-4})$ cm$^3$ s$^{-1}$, suggesting
    possible significant formation pumping effects in high-$A_V$ translucent
    clouds (see Appendix~\ref{appb}).} 
\begin{equation}
k_{\mathrm{col}} (T_{\mathrm{gas}})< 8\times10^{-11}
\times
\frac{\epsilon_{\mathrm{H_3O^+}}}{0.4} 
\times
\frac{\zeta/n_{\mathrm{H}}}{10^{-17}\,\mathrm{cm^3\, s^{-1}}}
\times
\frac{5\times10^{-8}}{\chi(\mathrm{H_3O^+})},
\end{equation}
where $k_{\mathrm{col}}$ (in cm$^3$ s$^{-1}$) is the rate for collisional
relaxation from a metastable level at \tgas, $\zeta$ is the ionization rate
per H nucleus, $\chi$ denotes the abundance relative to H nuclei, and
the reference value of $\epsilon_{\mathrm{H_3O^+}}=0.4$ is 
based on the chemical models of Sect.~\ref{sec:chem}. 
Collisional rates for \htop\ excitation with H$_2$ have been calculated by 
\cite{off92}\footnote{Available at
  http://www.strw.leidenuniv.nl/$\sim$moldata/} for a temperature of 100 K and
for the lowest \htop\ levels in the $K=0,1,2,3$ ladders. Taking as a guide
$k_{\mathrm{col}}\sim2\times10^{-10}$ cm$^3$ s$^{-1}$ at 100 K from the
$3_3^+$ level \citep{off92}, formation pumping may have important 
effects at high $\zeta/n_{\mathrm{H}}$, but collisional rates for higher
levels and with H$_2$ and H are required to check this
point. Calculations for excitation of \htop\ through electron impact
  \citep{fau03} indicate that the $\Delta K>0$ transitions are negligible, so
  collisions with electrons are probably not important in populating the
  $K>1$ metastable levels.

In our calculations for \htop\ below, we use the de-excitation rates by
\cite{off92}, and correct the excitation rates according to the value of
\tgas\ in our models and the requirements of detailed balance. Since no
collisional rates that connect the low- and high-$K$ ladders are available,
we simply force the metastable high-lying $J_{J^+}$ levels to
be populated according to the adopted \trot, which represents \tgas, the
signature of the formation process, or a combination of both. This will be
further discussed in Sect.~\ref{sec:discussion}. Additionally, we
attempt to establish in Appendix~\ref{appb} under which conditions the
formation process can generate the observed high excitation in environments
with moderate \tgas. Excitation of non-metastable levels is 
dominated by the far-IR field, and significant discrepancies between
observations and model results would indicate the relevance of alternative
excitation mechanism.

\subsection{Source components}

As shown in Paper I, the far-IR spectra of NGC 4418 and Arp 220 cannot be
described by a single set of ISM parameters, but different lines have
different excitation requirements and are thus formed in different regions of
the galaxies associated with different far-IR continuum
components that we summarize here (see also Figures 1 \& 16 and Tables 1
  \& 2 in Paper I). For NGC~4418, at far-IR wavelengths we require
$(i)$ a nuclear core
($C_{\mathrm{core}}$) with $T_{\mathrm{dust}}\approx130-150$ K, that
provides absorption in the high-lying lines of \hdo, OH, HCN, and NH$_3$;
$(ii)$ a warm component ($C_{\mathrm{warm}}$) with
$T_{\mathrm{dust}}\approx110-85$ K, that provides absorption in moderately
excited lines of \hdo\ and OH as well as a significant fraction of the far-IR
continuum emission; and $(iii)$ an extended, possibly infalling component
($C_{\mathrm{extended}}$) with $T_{\mathrm{dust}}\approx90-40$ K, which accounts
for low-lying redshifted lines of OH and O$^0$. For Arp 220, we require
$(i)$ the western and eastern nuclear components
($C_{\mathrm{west}}$ and $C_{\mathrm{east}}$), where the high-lying molecular
lines are formed; $(ii)$ the extended component ($C_{\mathrm{extended}}$),
which provides a significant fraction of the far-IR continuum emission (though
mostly reemission from the nuclei) and accounts for the absorption in
moderate and low-excitation lines of \hdo\ and OH; and $(iii)$ an additional
low-excitation, absorbing component or ``halo'' ($C_{\mathrm{halo}}$), with no
associated continuum that is required to fit the absorption in the
ground-state lines of \hdo, OH, and O$^0$. 

Calculations of excitation, line fluxes, and line profiles were first
performed for the nuclear components in both sources to fit the high-lying
lines, and then were done for the extended components to complete 
fitting of the absorption in the low-lying lines. Table~\ref{tab:columns} 
gives the inferred column densities, based on a ``screen'' approach for
the nuclear components (i.e. the molecules are located in front of the nucleus)
due to the high continuum opacities of these components in the far-IR;
the ``mixed'' approach (i.e. the molecules and dust are mixed) would give
columns highly exceeding the listed values \citep{gon08}. For the
$C_{\mathrm{extended}}$, the ``mixed'' approach is used 
owing to the much more moderate continuum opacities (Paper I).

   \begin{table*}
      \caption[]{Line equivalent widths and absorption fluxes.}
         \label{tab:fluxes}
          \begin{tabular}{lccccccc}   
            \hline
            \noalign{\smallskip}
  & & & & \multicolumn{2}{c}{NGC 4418} & \multicolumn{2}{c}{Arp 220} \\
Line  &  $E_{\mathrm{lower}}$ & $\lambda$ & $A_{ul}$ &
$W_{\mathrm{eq}}^{\mathrm{a}}$ & Flux$^{\mathrm{a}}$ & $W_{\mathrm{eq}}^{\mathrm{a}}$ & Flux$^{\mathrm{a}}$ \\  
  & (K) & ($\mu$m)  & (s$^{-1}$) & ($\mathrm{km\,\,s^{-1}}$) & 
($10^{-21}$ W cm$^{-2}$) & ($\mathrm{km\,\,s^{-1}}$) & ($10^{-21}$ W cm$^{-2}$) \\
            \noalign{\smallskip}
            \hline
            \noalign{\smallskip}
$\mathrm{OH^+} \, 2_1-1_0$ & $43.6$ & $148.696$ & $0.10$  & $6.7(1.8)$ &$0.79(0.21)$ & $17.8(0.9)$ & $10.0(0.5)$  \\
$\mathrm{OH^+} \, 2_3-1_2$ & $46.6$ & $152.989$ & $0.18$  & $^{\mathrm{b}}$ & $^{\mathrm{b}}$ & $^{\mathrm{b}}$ & $^{\mathrm{b}}$ \\
$\mathrm{OH^+} \, 2_2-1_2$ & $46.6$ & $147.768$ & $0.048$  & $3.9(1.6)$ &$0.47(0.19)$ & $20.7(1.2)$ & $12.0(0.7)$ \\
$\mathrm{OH^+} \, 2_2-1_1$ & $49.6$ & $152.369$ & $0.14$  & $19.5(2.1)$ &$2.1(0.2)$ & $45.3(1.3)$ & $24.6(0.7)$ \\
$\mathrm{OH^+} \, 2_1-1_1$ & $49.6$ & $158.437$ & $0.071$ & $5.9(3.0)$ & $0.6(0.3)$& $19.8(1.1)$ &  $9.6(0.5)$ \\
$\mathrm{OH^+} \, 4_5-3_4$ & $281.9$ & $76.510$ & $1.6$  & $3.1(0.7)$ &$1.6(0.4)$ & $5.8(0.4)$ & $10.6(0.7)$  \\
$\mathrm{OH^+} \, 4_3-3_2$ & $282.5$ & $76.245$ & $1.5$  & $3.2(0.8)$ &$1.6(0.4)$ & $3.1(0.4)$ & $5.7(0.7)$  \\
$\mathrm{OH^+} \, 4_4-3_3$ & $285.5$ & $76.399$ & $1.5$  & $3.8(0.8)$ &$1.9(0.4)$ & $4.0(0.4)$ & $7.4(0.7)$  \\
$\mathrm{pH_2O^+} \, 2_{12}\frac{5}{2}-1_{01}\frac{3}{2}$ & $0.0$ & $184.339$ & $0.086$ & $<4.8$ &$<0.25$ & $47.6(2.1)$ & $10.5(0.5)$\\ 
$\mathrm{pH_2O^+} \, 2_{12}\frac{3}{2}-1_{01}\frac{1}{2}$ & $0.1$ & $182.915$ & $0.073$ & $<3.7$ &$<0.21$ & $30.8(2.1)$ & $7.5(0.4)$ \\ 
$\mathrm{pH_2O^+} \, 2_{12}\frac{3}{2}-1_{01}\frac{3}{2}$ & $0.0$ & $182.640$ & $0.016$ & $<3.7$ &$<0.21$ & $14.5(1.6)$ & $3.6(0.4)$ \\ 
$\mathrm{pH_2O^+} \, 2_{21}\frac{5}{2}-1_{10}\frac{3}{2}$ & $29.1$ & $105.742$& $0.46$ & $7.6(2.3)$ &$2.1(0.6)$ & $6.2(0.8)$& $7.7(1.0)$ \\ 
$\mathrm{pH_2O^+} \, 2_{21}\frac{3}{2}-1_{10}\frac{1}{2}$ & $30.4$ & $104.719$& $0.39$ &$^{\mathrm{b}}$&$^{\mathrm{b}}$&$^{\mathrm{b}}$&$^{\mathrm{b}}$\\ 
$\mathrm{oH_2O^+} \, 3_{13}\frac{7}{2}-2_{02}\frac{5}{2}$ & $89.1$ & $143.813$ & $0.18$ & $5.2(2.5)$ &$0.65(0.31)$ &$22.6(1.1)$&$14.0(0.7)$\\ 
$\mathrm{oH_2O^+} \, 3_{13}\frac{5}{2}-2_{02}\frac{3}{2}$ & $89.3$ & $143.270$ & $0.17$ & $<3.0$ &$<0.38$  & $8.3(0.9)$& $5.2(0.6)$  \\ 
$\mathrm{oH_2O^+} \, 3_{22}\frac{7}{2}-2_{11}\frac{5}{2}$ & $125.0$ & $89.590$ & $0.59$ & $7.2(2.3)$ &$2.9(0.9)$ &$10.2(0.7)$&$16.6(1.1)$\\ 
$\mathrm{oH_2O^+} \, 3_{22}\frac{5}{2}-2_{11}\frac{3}{2}$ & $125.9$ & $88.978$ & $0.57$ & $5.4(1.6)$ &$2.2(0.6)$ & $2.2(1.1)$ &$3.6(1.8)$\\ 
$\mathrm{oH_2O^+} \, 4_{04}\frac{9}{2}-3_{13}\frac{7}{2}$ & $189.2$ & $145.917$& $0.16$ & $4.4(0.8)$ &$0.55(0.10)$ & $2.0(0.5)$ &$1.2(0.3)$\\ 
$\mathrm{oH_2O^+} \, 4_{04}\frac{7}{2}-3_{13}\frac{5}{2}$ & $189.8$ & $146.217$& $0.15$ & $4.7(0.8)$ &$0.58(0.10)$ & $5.5(0.6)$ &$3.2(0.4)$\\ 
$\mathrm{oH_3O^+} \, 3_{3}^--3_{3}^+$ & $102.0$ & $180.209$& $0.083$ &$<3.9$ &$<0.25$ & $28(3)$$^{\mathrm{c}}$ & $8.5(1.0)$$^{\mathrm{c}}$ \\ 
$\mathrm{pH_3O^+} \, 4_{4}^--4_{4}^+$ & $179.0$ & $178.994$& $0.091$ &$<3.9$ &$<0.27$ & $12.7(1.3)$ & $4.0(0.4)$ \\ 
$\mathrm{pH_3O^+} \, 5_{5}^--5_{5}^+$ & $273.4$ & $177.272$& $0.097$ &$<6.0$ &$<0.44$ & $13.9(1.6)$ & $4.4(0.5)$ \\ 
$\mathrm{oH_3O^+} \, 6_{6}^--6_{6}^+$ & $385.1$ & $175.063$& $0.10$ &$<4.6$ &$<0.38$ & $18.6(1.3)$ & $6.4(0.4)$ \\ 
$\mathrm{pH_3O^+} \, 7_{7}^--7_{7}^+$ & $514.0$ & $172.396$& $0.11$ &$<3.4$ &$<0.28$ & $11.8(1.4)$ & $4.2(0.5)$ \\ 
$\mathrm{pH_3O^+} \, 8_{8}^--8_{8}^+$ & $660.1$ & $169.303$& $0.12$ &$^{\mathrm{d}}$ &$^{\mathrm{d}}$ & $13.7(1.2)$ & $5.4(0.5)$ \\ 
$\mathrm{oH_3O^+} \, 9_{9}^--9_{9}^+$ & $823.3$ & $165.819$& $0.13$ &$8.0(2.0)$ &$0.71(0.18)$ & $16.5(2.0)$$^{\mathrm{e}}$&$6.9(0.8)$$^{\mathrm{e}}$\\ 
$\mathrm{pH_3O^+} \, 10_{10}^--10_{10}^+$ & $1003.5$ & $161.984$& $0.14$ &$<3.7$ &$<0.36$ & $7.5(1.1)$ & $3.3(0.5)$ \\ 
$\mathrm{oH_3O^+} \, 12_{12}^--12_{12}^+$ & $1414.7$ & $153.424$& $0.17$ & & & $8.5(1.5)$$^{\mathrm{f}}$ & $4.5(0.8)$$^{\mathrm{f}}$\\ 
$\mathrm{pH_3O^+} \, 5_{4}^--4_{4}^+$ & $179.0$ & $60.164$& $0.49$ &$5.8(2.0)$ &$4.1(1.4)$ & $5.5(1.2)$ & $12.1(2.7)$ \\ 
$\mathrm{oH_3O^+} \, 4_{3}^--3_{3}^+$ & $102.0$ & $69.538$& $0.38$ &$6.1(1.3)$ &$3.7(0.8)$ & $17.0(1.5)$ & $33.6(2.9)$ \\ 
$\mathrm{pH_3O^+} \, 4_{2}^--3_{2}^+$ & $139.3$ & $70.254$& $0.63$ &$<2.3$ &$<1.4$ & $5.4(0.7)$ & $10.4(1.4)$ \\ 
$\mathrm{pH_3O^+} \, 4_{1}^--3_{1}^+$ & $161.7$ & $70.684$& $0.77$ &$5.9(1.9)$ &$3.5(1.1)$ & $4.8(0.5)$ & $9.1(1.0)$ \\ 
$\mathrm{oH_3O^+} \, 4_{0}^--3_{0}^+$ & $169.1$ & $70.827$& $0.81$ &$8.0(2.0)$ &$4.6(1.2)$ & $7.7(0.8)$ & $15.8(1.5)$ \\ 
$\mathrm{pH_3O^+} \, 3_{2}^--2_{2}^+$ & $42.3$ & $82.274$& $0.28$ &$4.0(1.3)$ &$1.8(0.6)$ & $7.9(0.6)$ & $13.8(1.0)$ \\ 
$\mathrm{pH_3O^+} \, 3_{1}^--2_{1}^+$ & $64.7$ & $82.868$& $0.44$ &$<1.2$ &$<0.6$ & $13.1(0.7)$ & $23.2(1.2)$ \\ 
     \noalign{\smallskip}
  \hline
         \end{tabular} 
\begin{list}{}{}
\item[$^{\mathrm{a}}$] Numbers in parenthesis indicate $1\sigma$ uncertainties. 
\item[$^{\mathrm{b}}$] The line is strongly contaminated.
\item[$^{\mathrm{c}}$] Contaminated by \hdo\ and CH (Fig.~\ref{spectrah3o+}a).
\item[$^{\mathrm{d}}$] Dominated by HCN (Paper I). 
\item[$^{\mathrm{e}}$] Contaminated by \nht; the line flux in Arp~220 is
  estimaded by subtracting the modeled \nht\ flux from the flux of the whole
  feature (Fig.~\ref{spectrah3o+}g). 
\item[$^{\mathrm{f}}$] Tentative detection, at the wing of the NH 153 $\mu$m
  line (Figs.~\ref{spectraoh+}a and \ref{spectrah3o+}i).
\end{list}
   \end{table*}

\subsection{NGC 4418}

\subsubsection{OH$^+$}

In NGC~4418 we first used the parameters of the \ccore\
and \cwarm\ components as given in Paper I to match the absorption in the
high-lying $4_{J}\leftarrow3_{J'}$ lines of \ohp, by assuming a covering
factor of 1. The \ohp\ column densities were similar for both components,
$N(\mathrm{OH^+})=(0.6-1)\times10^{16}$ \cmd, respectively. The model fits
were also very similar for both components, so that we are not able to
determine in which nuclear component the \ohp\ lines are
(primarily) formed. By using the same normalization column as in Paper I,
$N(\mathrm{H})=4\times10^{23}$ \cmd, the \ohp\ abundance in the nuclear region
of NGC~4418 is $\chi(\mathrm{OH^+})=(1.5-2.5)\times10^{-8}$.
Contrary to observed, the $4_{5}\leftarrow3_{4}$ component in
Fig.~\ref{spectraoh+}e is predicted to be the strongest, which could indicate
significant departures of the fine-structure distribution of populations
relative to the predicted one. The effective spatial extent (diameter)
predicted by the \ccore\ and \cwarm\ models is $\approx20-30$ pc,
respectively, and $T_{\mathrm{dust}}=150-110$ K.  

Figure~\ref{spectraoh+} shows (blue curves) the model fit for the 
nuclear component (similar for \ccore\ and \cwarm),
together with a model for NH in panel a (light-blue curve).
The nuclear model that fits the $4_{J}\leftarrow3_{J'}$ lines underpredicts the
$2_{2}\leftarrow1_{1}$ line and the $153.0$ $\mu$m feature (composed of the
\ohp\ $2_{3}\leftarrow1_{2}$ and the NH $2_{2}\leftarrow1_{1}$ lines, see
Fig.~\ref{spectraoh+}a). The \ohp\ $2_{2}-1_{1}$ line has a redshifted excess,
and the $153.0$ $\mu$m feature cannot be fully reproduced at just the central 
wavelength of the feature, which is also redshifted relative to the \ohp\
line. This strongly suggests that the absorption that cannot be fit
  with the nuclear model is due to \cext, a more extended,
  inflowing component proposed in Paper I in order to explain the redshifted
  absorption observed in the ground-state lines of OH and [O {\sc i}]. 
Fitting the remaining \ohp\ absorption with the \cext\ model in Paper I
(green curves in Figure 2a-upper), the fit shown with the red curve in Figure
2a is obtained. We infer $N(\mathrm{OH^+})\sim3\times10^{15}$ \cmd, and
roughly $\chi(\mathrm{OH^+})\sim2\times10^{-8}$. 

In summary, the \ohp\ lines are essentially formed in the nuclear
$C_{\mathrm{warm}}+C_{\mathrm{core}}$ components, with 
$N(\mathrm{OH^+})\approx(6-10)\times10^{15}$ \cmd. Some contribution 
to the $2_{2}\leftarrow1_{1}$ and $2_{3}\leftarrow1_{2}$ lines
due to inflowing gas in \cext\ is also inferred, with
$N(\mathrm{OH^+})\sim3\times10^{15}$ \cmd.

\subsubsection{H$_2$O$^+$}

H$_2$O$^+$ is weak in NGC~4418 (Sect.~\ref{obs:h2o+}),
though the $5/2\leftarrow3/2$ component of the \t322211\ transition at 89
$\mu$m, and more tentatively the $7/2\leftarrow5/2$ one, are detected in the
source (Fig.~\ref{spectrah2o+}d). We have carried out models for the
$C_{\mathrm{core}}$ and $C_{\mathrm{warm}}$ components that fit the above
lines. Both models give similar columns of
$N(\mathrm{H_2O^+})\approx3-4\times10^{15}$ \cmd, but fail
to account for the somewhat marginal \t404313\ components
(Fig.~\ref{spectrah2o+}e). 
The \ccore\ model prediction better matches these features but, as
  shown with blue curves in Fig.~\ref{spectrah2o+}c, may (marginally) 
overestimate the \t313202\ components. The \hdop\ \t331220\
$7/2\leftarrow5/2$ at $65.6$ $\mu$m (panel f) is blended with NH$_2$ and
$^{18}$OH lines and affected by an uncertain baseline (Paper I). We conclude
that the \ohp/\hdop\ column density ratio in the nuclear region of NGC 4418 is
in the range $1.5-2.5$. The ground-state \hdop\ lines are not
detected, indicating that \hdop\ has a low abundance in the
\cext\ component of NGC 4418.

\subsubsection{H$_3$O$^+$}

Only the RI lines of \htop\ were detected in NGC 4418.  They
were fitted with the \ccore\ and \cwarm\
components, with an assumed ortho-to-para ratio of 1. Both components give
similar column densities of $N(\mathrm{H_3O^+})\approx5-8\times10^{15}$
\cmd, similar to that inferred by \cite{aal11} from observations of the
$3_2^+\leftarrow2_2^-$ line. The \ccore\ model is shown with blue curves in
Fig.~\ref{spectrah3o+}. The main drawback of both models is that the
absorption in the $3_1^-\leftarrow2_1^+$ transition (panel p) is significantly
overpredicted; the line is not detected, possibly due to cancellation
between absorption toward the continuum source and reemission from a more
extended region. The \ohp/\htop\ column density ratio in the nuclear region of
NGC~4418 is $\approx1$. 

The PIMS lines of \htop\ are not detected in NGC~4418, but the model for
\ccore\ in Fig.~\ref{spectrah3o+} uses the same $T_{\mathrm{rot}}=500$ K as
in Arp~220, and the predicted faint PIMS lines are nevertheless quite
compatible with the observed spectrum. The model predicts wing-like features
at the wavelengths of the $9_9^--9_9^+$ and $12_{12}^--12_{12}^+$ ortho PIMS
lines, which could be present in the spectrum at very low (but not
  significant) signal-to-noise. Therefore, the 
non-detection of the PIMS lines does not imply that \trot\ is low in NGC~4418,
but according to our model can also be due to moderate continuum opacity
and dilution of the nuclear continuum emission at long far-IR wavelengths.

   \begin{table*}
      \caption[]{Inferred column densities, \ohp\ abundances, and \ohp/\hdop\
        and \ohp/\htop\ column density ratios.}
         \label{tab:columns}
          \begin{tabular}{lcccc}   
            \hline
            \noalign{\smallskip}
            Source & \multicolumn{2}{c}{NGC 4418}  & \multicolumn{2}{c}{Arp 220}  \\
            Component & \ccore-\cwarm\ & \cext\ & \cwest\ & \cext\  \\ \hline
\tdust\ (K)  & $110-150$ & variable$^{\mathrm{c}}$  & $90-110$ & variable$^{\mathrm{c}}$ \\
$N$(\ohp)$^{\mathrm{a}}$ ($10^{16}$ cm$^{-2}$) & $0.6-1$ & $\sim0.3$ & $0.5-1.1$ & $2-4$ \\
$N$(\hdop)$^{\mathrm{a}}$ ($10^{16}$ cm$^{-2}$) & $0.3-0.4$ & & $0.25-0.9$ & $\sim0.5$$^{\mathrm{d}}$ \\
$N$(\htop)$^{\mathrm{a}}$ ($10^{16}$ cm$^{-2}$) & $0.5-0.8$ & & $0.9-2.7$$^{\mathrm{e}}$ &  \\
$\chi$(\ohp)$^{\mathrm{b}}$ ($10^{-8}$)  & $1.5-2.5$ & $\sim2$ & $1.1-2.8$ & $\sim4$ \\
\ohp/\hdop\ & $1.5-2.5$ & & $1.0-2.0$ & $5-10$ \\
\ohp/\htop\ & $\sim1$   & & $\sim1$$^{\mathrm{e}}$ &  \\
               \noalign{\smallskip}
            \hline
         \end{tabular} 
\begin{list}{}{}
\item[$^{\mathrm{a}}$] Columns toward the nuclei are calculated
  within a screen approach.
\item[$^{\mathrm{b}}$] Estimated \ohp\ abundance relative to H nuclei, using a
  normalization column density of $N(\mathrm{H})=4\times10^{23}$ \cmd\ for the
  nuclear components (Paper I).  
\item[$^{\mathrm{c}}$] $T_{\mathrm{dust}}$ varies with radial position as it is calculated
  from the balance between heating and cooling. 
\item[$^{\mathrm{d}}$] The p-H$_2$O$^+$ ground-state lines at $183-184$ $\mu$m
  in Arp 220 require an additional absorbing halo component (\chalo) with
  $N(\mathrm{H_2O^+})\sim(2.5-3.0)\times10^{15}$ \cmd. 
\item[$^{\mathrm{e}}$] In Arp 220, the total \htop\ column density is twice the
  nuclear \ohp\ column, but since the \htop\ lines are redshifted relative to
  the \ohp\ lines, we roughly estimate an \ohp/\htop\ ratio of 1 in the
  nuclear region where the \ohp\ lines are formed. In the \cext\ component,
  the \ohp/\htop\ ratio is uncertain.
\end{list}
   \end{table*}

\subsection{Arp 220}
\label{sec:arp220models}

\subsubsection{OH$^+$}

In Arp 220, the high-lying \ohp\ $4_{J}\leftarrow3_{J'}$ lines 
shown in Fig.~\ref{spectraoh+}e are tentatively 
associated with the luminous western nucleus, $C_{\mathrm{west}}$, 
but they can be generally interpreted as the joint emission from 
the two nuclei (Paper I). In our model for $C_{\mathrm{west}}$, we
include a velocity gradient through the absorbing shell of 130
\kms, to reproduce the observed line broadening. 
We use two approaches to bracket the parameters inferred in Paper I
for \cwest. For a compact/warm region, with $T_{\mathrm{dust}}=110$ K
and $d=93$ pc, the high-lying \ohp\ $4_{J}\leftarrow3_{J'}$ lines are well
fitted with a column of $N(\mathrm{OH^+})=1.1\times10^{16}$ \cmd, similar to
the value found in NGC~4418. The predicted spectrum is shown with blue curves
in Fig.~\ref{spectraoh+}. The \ohp\ absorption in the $4_{J}\leftarrow3_{J'}$
lines, however, does not require such high $T_{\mathrm{dust}}$. For a
colder and more extended region, with $T_{\mathrm{dust}}=90$ K and
$d=160$ pc, the column density is $N(\mathrm{OH^+})=4.5\times10^{15}$ \cmd.
These models, however, do not reproduce the strong $2_{J}\leftarrow1_{J'}$ and
$3_{2}\leftarrow2_{2}$ lines (Fig.~\ref{spectraoh+}a-d), 
so another component is required in addition.
We identify this additional component as  
$C_{\mathrm{extended}}$ (Paper I), which has moderate excitation in
\hdo\ and OH. The \cext\ 
component can be identified with the (inner region of) the external disk
traced by CO \citep{dow07}. In Paper I we estimated $d\sim650$ pc, though this
is uncertain by at least a factor of 2. The additional \ohp\ absorption could
  indeed be produced in a relatively small region ($d\lesssim300$ pc) around
  the nuclei; our results are relatively uncertain for this component.
Assuming that the extended \ohp\ mainly absorbs the continuum of the
  \cext\ component (Fig.~1 in Paper I), it most probably has a column of
$N(\mathrm{OH^+})\approx(2-4)\times10^{16}$ \cmd.
This reflects the widespread presence of \ohp\ in the regions surrounding the
nuclei.

The joint contribution of $C_{\mathrm{west}}$ and $C_{\mathrm{extended}}$ to
the \ohp\ absorption, shown with a red curve in Fig.~\ref{spectraoh+}, 
yields a reasonable fit to most \ohp\ lines in Arp 220, though the
$3_{2}\leftarrow2_{2}$ line at 104 $\mu$m remains underpredicted 
(and blended with an \hdo\ line). The 
wing-like spectral feature at $153.45$ $\mu$m is not reproduced, though it is
probably due to \htop\ (Sect.~\ref{sec:mod:h3o+arp220}).

\subsubsection{H$_2$O$^+$}

The above models for \cwest\ were applied to \hdop. Using the parameters for
the compact/warm approach, $T_{\mathrm{dust}}=110$ K and $d=93$ pc, we derive
a \hdop\ column density similar to that found for \ohp,
$N(\mathrm{H_2O^+})\approx0.9\times10^{16}$ \cmd. The model result is shown in 
Fig.~\ref{spectrah2o+} (blue curves). 
In this model the \t322211\ $5/2\leftarrow3/2$ and possibly the \t331220\
  $7/2\leftarrow5/2$ lines are overpredicted, but the other ortho-\hdop\ lines
  are well reproduced.
Using the more extended, colder approach, with $T_{\mathrm{dust}}=90$ K
and $d=160$ pc, the column density is $N(\mathrm{H_2O^+})=2.5\times10^{15}$
\cmd. In the nuclear region of Arp~220, the most likely \ohp-to-\hdop\ ratio
is $1-2$. 

Apart from the para-\hdop\ ground transition, there is only one excited,
  non-contaminated para-\hdop\ line detected in the spectrum of Arp~220, the
  \t221110\ $5/2-3/2$ at 105.7 $\mu$m. As can be seen in
  Fig.~\ref{spectrah2o+}b, the absorbing flux in this line is well reproduced
  with our models, which use an ortho-to-para ratio of 3, the 
  high-temperature limit. This provides a slightly
  better fit than the one obtained with an ortho-to-para ratio of
  $\approx4.8$, the value derived by \cite{sch10} toward Sgr~B2, which
  underestimates the absorbing line flux in Arp~220 by 35\%.   

The ground-state p-\hdop\ \t212101\ lines at $183-184$ $\mu$m cannot be
reproduced with the \cwest\ component (Fig.~\ref{spectrah2o+}a). Models for
the \cext\ were also unsuccessful in reproducing these spectral features, as
the required high ($1.6\times10^{16}$ \cmd) \hdop\ columns have also the
effect of predicting too strong absorption in other lines, specifically in the
\t322211\ components. The p-\hdop\ 183-184 $\mu$m lines peak at a velocity
different from the \hdop\ HE lines (Fig.~\ref{vel}), and are thus likely
generated in a low excitation region that generates faint absorption in other
lines. This component is likely to be the same as that producing strong
absorption in the ground-state lines of \hdo\ and OH, i.e. the \chalo\
component (Paper I). The spatial extent of this component is not constrained,
and can be interpreted as foreground gas located along the line of sight to
the nuclei absorbing their nuclear far-IR emission. As shown below
(Sect.~\ref{sec:mod:h3o+arp220}), however, it probably includes gas within a
few hundreds pc from the nuclei; the \cext\ and \chalo\ components are hard to
separate physically\footnote{The difference between the \cext\ and
  \chalo\ components in our models for Arp~220 arises from radiative transfer
  effects. The \cext\ is characterized by optically thin, extended far-IR
  emission, where the molecules are still significantly excited via absorption
  of locally-emitted far-IR photons. The \chalo, however, describes
  foreground, low-excitation gas with strong, optically thick emission behind
  arising from the nuclei. The main difference may simply be the location of
  the molecules relative to the far-IR optically thick nuclei and the
  observer, thus basically describing the same spatially extended gas.}. 
The derived \hdop\ column in this component is
$N(\mathrm{H_2O^+})\approx3\times10^{15}$ \cmd, and the fit is shown with a
light blue curve in Fig.~\ref{spectrah2o+}a.

   \begin{figure}
   \centering
   \includegraphics[width=8.5cm]{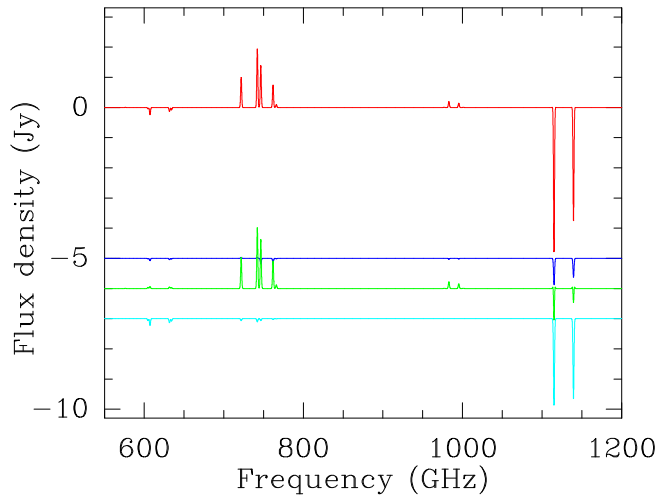}
   \caption{Predicted SPIRE spectrum of H$_2$O$^+$ in Arp 220 between 600 and
     1200 GHz. The contributions by the \cwest, \cext, and \chalo\ are shown
     with blue, green, and light-blue curves (vertically shifted for clarity),
     respectively, and the red curve shows the total absorption/emission. }  
    \label{h2o+_spire}
    \end{figure}

The amount of H$_2$O$^+$ in $C_{\mathrm{extended}}$ is better traced with
the excited \hdop\ detected in emission with Herschel/SPIRE by R11. In
Fig.~\ref{h2o+_spire} we show the predicted SPIRE spectrum of \hdop\ in Arp
220 between 600 and 1200 GHz. The contributions by the \cwest, \cext, and
\chalo\ are shown with blue, green, and light-blue curves (vertically shifted
for clarity), respectively, and the red curve shows the total
absorption/emission. 
At these frequencies, the contribution by the nuclear region (\cwest) is
not expected to be dominant. The absorption in the ground-state \t111000\
$J=3/2\leftarrow1/2$ and $1/2\leftarrow1/2$ o-\hdop\ lines at $1100-1150$
GHz is expected to be dominated by the \chalo\ component, also responsible for
the absorption in the ground-state p-\hdop\ lines at $183-184$ $\mu$m.
Most relevant is the predicted {\em emission} in some \hdop\ lines at $700-800$
GHz, which are expected to arise from the \cext\ component. The strongest
\hdop\ emission lines are the $2_{02}\rightarrow1_{11}$ $5/2\rightarrow3/2$ at
$742.1$ GHz and the $2_{11}\rightarrow2_{02}$ $5/2\rightarrow5/2$ at $746.5$
GHz, in agreement with the emission features reported by R11\footnote{R11
  assigned the $746.5$ GHz feature to the $2_{02}\rightarrow1_{11}$
  $3/2\rightarrow3/2$ at nearly the same frequency, 
  but this transition has an Einstein coefficient 5.4 lower than the
  main $5/2\rightarrow3/2$ component and is expected to give weak emission}.
Two other transitions with significant predicted emission above the
continuum are the $2_{02}\rightarrow1_{11}$ $3/2\rightarrow1/2$ at $721.8$ GHz
and the $2_{11}\rightarrow2_{02}$ $3/2\rightarrow3/2$ at $761.9$ GHz. 
It is worth noting that the modeled emission features are generated through
radiative pumping and not through collisions -collisional excitation is not
included in our calculations. With a column density in \cext\ of
$N(\mathrm{H_2O^+})\approx5\times10^{15}$ \cmd, the lines fluxes at 
  $700-800$ GHz reported by R11 are nearly reproduced.
Our models thus favor an \ohp/\hdop\ column density ratio of $\gtrsim5$ 
for \cext, significantly higher than in the nuclear region
(Table~\ref{tab:columns}). In the \chalo, the \ohp/\hdop\ is uncertain as the
\ohp\ ground-state lines are certainly saturated (R11).

\subsubsection{H$_3$O$^+$}
\label{sec:mod:h3o+arp220}

The \htop\ lines in Arp~220 show an intriguing behavior: on the one hand,
non-metastable RI lines are detected (Fig.~\ref{spectrah3o+}lmnp); on the
other hand, although they are radiatively pumped by far-IR photons to
  moderate excitation, 
these lines peak at roughly the same velocity as the LE lines of \hdo\ and
\hdop\ transitions (Fig.~\ref{vel}), i.e. they are redshifted relative to the
HE lines of other species. 

From the population diagram in Fig.~\ref{diagh3o+}, we have used two values
for \trot: $180$ K up to the $K=5$ ladder, and $500$ K for higher-lying
metastable levels. The blue lines in Fig.~\ref{spectrah3o+} show the model
results with $T_{\mathrm{dust}}=90$ K, $d=160$ pc, and
$N(\mathrm{H_3O^+})=9\times10^{15}$ \cmd. The model for \htop\ nearly accounts
for the pure-inversion metastable lines (panels a-i), which have moderate
opacities ($\tau\sim0.3-1$), though the $3_{3}^{-}\leftarrow3_{3}^{+}$ line is
slightly underpredicted. The rotation-inversion lines (panels j-p) are more
optically thin.

\begin{figure}
   \centering
   \includegraphics[width=8.5cm]{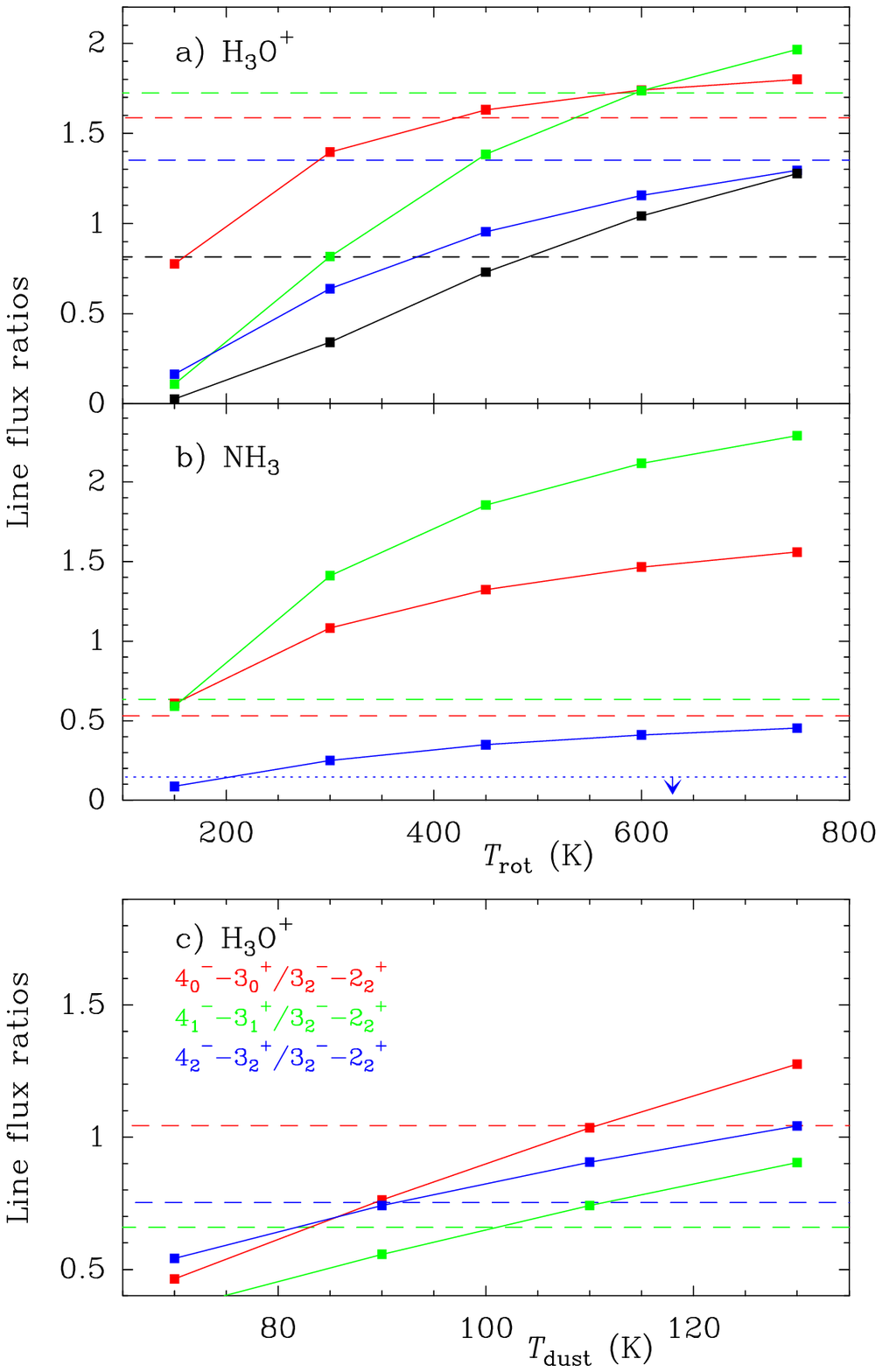}
\caption{Model results for \htop\ and \nht\ in Arp~220. a) Ratios of
pure inversion metastable lines of \htop\ versus \trot: $K=6/K=4$ (red),
$K=8/K=4$ (blue), $K=9/K=4$ (green), and $K=10/K=4$ (black). b) For \nht, the
fluxes of the $6_5^--5_5^+$ (red), $7_6^+-6_6^-$ (green), and $8_6^--7_6^+$
(blue) are normalized to the flux of the $124.9$ $\mu$m feature (composed of
the $4_0-3_0$ and $4_1^--3_1^+$ \nht\ lines, Paper I). c) Ratios
of rotation-inversion lines of \htop\ as a function of \tdust. 
Dashed lines in all panels indicate the
measured values, with typical uncertainties of 20\%.    
}
\label{nh3h2o}
\end{figure}

Figure~\ref{nh3h2o}a shows the flux ratios of the pure inversion,
metastable $6_{6}^{-}\leftarrow6_{6}^{+}$, $8_{8}^{-}\leftarrow8_{8}^{+}$,
$9_{9}^{-}\leftarrow9_{9}^{+}$, and $10_{10}^{-}\leftarrow10_{10}^{+}$ lines to the
$4_{4}^{-}\leftarrow4_{4}^{+}$ line (solid lines) as a function of \trot, and
compares them with the measured values (dashed lines). The flux measured for the
$9_{9}^{-}\leftarrow9_{9}^{+}$ line has been estimated from the total flux of
the broad $165.8$ $\mu$m feature by subtracting the modeled contribution due
to \nht\ (Fig.~\ref{spectrah3o+}g). The $K=8/K=4$ (blue) and $K=9/K=4$ 
(green) ratios indicate $T_{\mathrm{rot}}\sim600-800$ K, while the 
$K=6/K=4$ (red) and $K=10/K=4$ (black) favor
$\sim450$ K. The high \trot\ inferred from the \htop\ lines is in striking
contrast with that inferred from \nht\ (Fig.~\ref{nh3h2o}b): the \nht\ line
ratios indicate $T_{\mathrm{rot}}\sim 150$ K, in general agreement with the value
inferred by \cite{ott11} from the pure inversion lines, and with the HCN
excitation (Paper I). Either the two species are sampling different regions,
and/or \trot\ as derived from \htop\ better reflects the formation
process rather than \tgas.   

The strength of the non-metastable rotation-inversion lines is
sensitive to \tdust, and in Fig.~\ref{nh3h2o}c we compare the fluxes of
some of these lines relative to that of the metastable
$3_{2}^{-}\leftarrow2_{2}^{+}$ line. We focus in this analysis on those lines
that have similar line shapes, i.e. the $4_{0}^{-}\leftarrow3_{0}^{+}$,
$4_{1}^{-}\leftarrow3_{1}^{+}$, and $4_{2}^{-}\leftarrow3_{2}^{+}$
transitions. Comparison with the measured values favors 
$T_{\mathrm{dust}}\sim 90-110$ K. The most reliable comparison is
between the $4_{2}^{-}\leftarrow3_{2}^{+}$ and the
$3_{2}^{-}\leftarrow2_{2}^{+}$, as they belong to the same K-ladder, favoring
$T_{\mathrm{dust}}\sim 90$ K. Furthermore, due to the warmer SED,
$T_{\mathrm{dust}}=110$ K generally underestimates the PIMS lines relative to
the RI lines. In contrast, the \hdo\ HE lines are compatible with higher
\tdust\ (Paper I), and detection of the \nht\ $6_{0}^{-}\leftarrow5_{0}^{+}$ and
$6_{1}^{-}\leftarrow5_{1}^{+}$ lines at $83.85$ $\mu$m (Paper I and confirmed
with the new data set), with $E_{\mathrm{low}}\sim400$ K and pumped through
radiation, better supports $T_{\mathrm{dust}}=110-130$ K for \nht. 

The value of $N(\mathrm{H_3O^+})$ depends on the size of the far-IR
continuum source. To produce the observed \htop\ absorptions, the models
strongly favor optically thick far-IR continuum emission up to at least 180
$\mu$m. An upper limit to the size would then be $230$ pc, as the continuum
source with $T_{\mathrm{dust}}= 90$ K would produce the entire bolometric
luminosity of the galaxy, $1.5\times10^{12}$ \Lsun. This upper limit is,
however, unreliable, as \cext\ is expected to generate some far-IR emission,
and there is also an important fraction of the luminosity associated with dust
warmer than 90 K. Sizes smaller than $100$ pc are also unlikely, as the \htop\
PIMS lines would become optically thick, and the fit to the whole \htop\ SLED
would get worse (specifically the $6_6^-\leftarrow6_6^+$ line would be
underpredicted due to saturation). We estimate $d=105-160$ pc, overlapping
with, but at the high end of our derived size range for the nuclear \ohp\ and
\hdop\ ($93-160$ pc) and significantly smaller than our estimate for the
extended region as derived from the continuum emission ($\sim650$ pc, Paper
I).  This size range implies a column density range of  
$N(\mathrm{H_3O^+})=(2.7-0.9)\times10^{16}$ \cmd. These columns are a factor
of 2 higher than those of \ohp\ for the same source sizes. However, the \htop\
transitions do not have the same line shapes as the \ohp\ lines
(Fig.~\ref{vel}h). There is, nevertheless, a good match
between the \ohp\ HE and \htop\ 70.8 $\mu$m profiles on the blueshifted 
side, suggesting some spatial overlap between both
components. On these grounds, we favor a column density ratio
$\mathrm{OH^+/H_3O^+} \sim 1$ in the nuclear region of Arp 220, with an
uncertainty of $\pm50$\%. 
The relatively large size found for the \htop\ absorption, the shift of the
peak of the \htop\ velocity profile towards those of the low excitation lines,
and the moderate \tdust\ found for the \htop\ lines, suggest that the \htop\
lines, while nuclear, trace a transition region between the high-excitation
(HE) component and the extended component.

The model in Fig.~\ref{spectrah3o+} predicts a flux for the
$3_{2}^{+}\rightarrow2_{2}^{-}$ 365 GHz line of $1.1\times10^{-23}$ W \cmd, a
factor of $1.5$ lower than measured by vdT08. The modeled line is slightly
inverted but with negligible amplification, and thus behaves as optically thin
with the flux proportional to the column. Either it is sampling in addition
deep regions that are extincted in the far-IR, or it is sampling a larger
region (vdT08), because the 365 GHz emission line does not require
far-IR optically thick continuum emission behind.

\subsection{Summary}

The analysis of the O-bearing molecular ions in NGC~4418 and Arp~220 reveal
$(i)$ high nuclear \ohp\ column densities of $(0.5-1)\times10^{16}$ \cmd\ in
both sources, with estimated abundances relative to H nuclei of
$(1-3)\times10^{-8}$; 
$(ii)$ \ohp/\hdop\ column density ratios of $\sim1-2.5$, with NGC~4418 at
  the high end of the range; 
$(iii)$ high \ohp\ columns of $\sim(2.5-4)\times10^{16}$ \cmd\ in the \cext\
component of Arp 220, with an estimated abundance of $\sim4\times10^{-8}$, as
well as a high \ohp/\hdop\ ratio of $\sim5-10$; 
$(iv)$ \htop\ column densities of $\sim(0.5-0.8)\times10^{16}$ \cmd\ in
NGC~4418 and $\sim(0.9-2.7)\times10^{16}$ \cmd\ in Arp~220, the latter
essentially nuclear but likely tracing a region slightly more extended
than that traced by \ohp. In the nuclear region where the \ohp\
lines are formed, $\mathrm{OH^+/H_3O^+} \sim 1$ is estimated.

\section{Chemistry}
\label{sec:chem}

\subsection{Framework}

The O-bearing molecular ions are expected to be formed in gas irradiated by
cosmic rays or X-rays \citep{mal96,ger10,neu10,mei11}. The sequence is
initiated by the ionization of H atoms and H$_2$ molecules; \ohp\ and \hdop\
are subsequently produced by ion-neutral chemistry, and their abundances are
sensitive to the gas ionization rate $\zeta$ and the molecular fraction
$f_{\mathrm{H_2}}$. In regions with a significant atomic fraction, 
H$^+$ transfers its charge to atomic oxygen, and the O$^+$
reacts with H$_2$ to form \ohp; \hdop\ is then formed from
$\mathrm{OH^++H_2}$, and similarly \htop\ from $\mathrm{H_2O^++H_2}$
\citep{ger10,neu10}. In regions with high molecular fraction, \ohp\ and \hdop\
can also be formed via $\mathrm{H}_3^++\mathrm{O}$
\citep[e.g.][]{her73,ger10,hol12}. Furthermore, in environments with high
reservoirs of OH and \hdo, \ohp\ and \hdop\ can be 
produced from the charge exchange reactions $\mathrm{OH+H^+}$ and
$\mathrm{H_2O+H^+}$, as well as from photoionization of OH and \hdo\
\citep{gup10}, while \htop\ may also be formed from $\mathrm{H_2O+H_3^+}$.  
Finally, H$^+$ can also be produced through $\mathrm{C^++OH}$ and
$\mathrm{CO^++H}$, and this FUV/chemical route does not require X/cosmic
ray irradiation \citep{ste95,hol12}.

   \begin{figure}
   \centering
   \includegraphics[width=8.5cm]{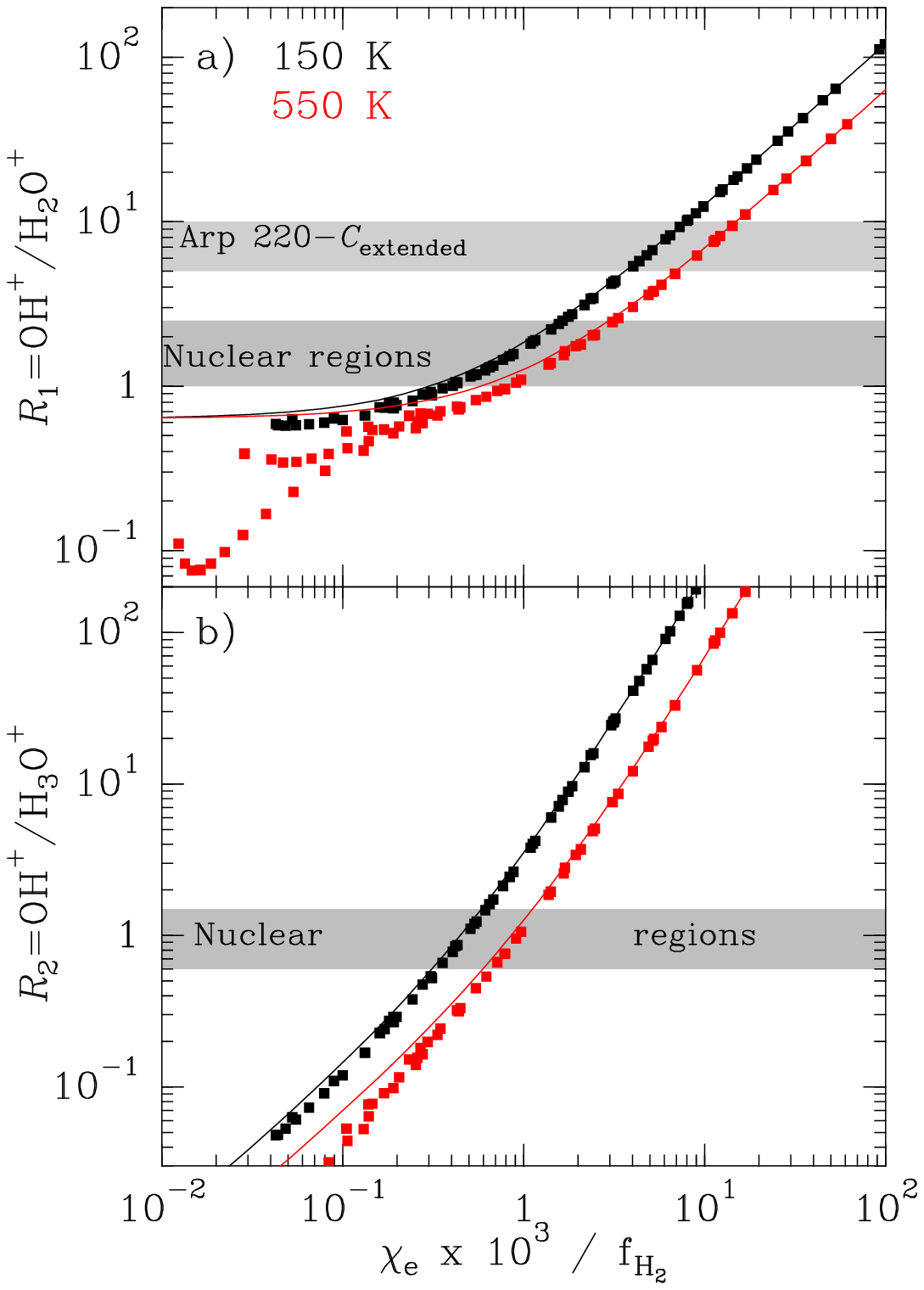}
   \caption{Predicted (a) \ohp/\hdop\ and (b) \ohp/\htop\ ratios as a function
     of $\chi_e\times10^3/f_{\mathrm{H_2}}$. Symbols show results for all
     generated chemical models, while the solid lines show the analytic
     curves by \cite{ger10} and \cite{neu10}. The inferred \ohp/\hdop\
     ratios in the nuclear regions of NGC 4418 and Arp 220, and in the \cext\
     of Arp 220, are indicated with the shaded dark and light regions,
     respectively. Gas temperatures of 150 K (black) and 550 K (red) are
     assumed.}   
    \label{ratio_oh+_h2o+}
    \end{figure}

In order to discriminate between the above paths of \ohp, \hdop, and \htop\
production, and to infer the physical conditions implied by the observations,
we have run a grid of chemical models with the code described by
\cite{bru09}, based on previous works by \cite{sta05} and \cite{dot02}. 
The chemical models calculate the steady state abundances of 
all relevant species based on the UMIST06 reaction rates \citep{woo07}. In our
simple models, the gas with H nuclei density $n_{\mathrm{H}}$ and molecular
fraction $f_{\mathrm{H_2}}$ is directly exposed to a cosmic ray and/or X-ray
flux that produces a total ionization rate per H nucleus $\zeta$ (including
secondary ionizations). We ignore any other external agent 
(e.g. dissociating UV radiation) 
in the calculations, though internally generated UV radiation is
included. 

In their models for the production of the O-bearing molecular ions in
  interstellar clouds, \cite{hol12} found that PAHs have important 
  effects on their expected abundances. Specifically, a fraction of H$^+$ ions
  produced by cosmic rays was found to be neutralized mostly by 
  PAH$^-$, truncating the path for the production of \ohp, and thus
  reducing the efficiency $\epsilon_{\mathrm{OH^+}}$ of \ohp\ formation
  relative to previous calculations \citep{neu10}. Recent observations of
  both \ohp\ and H$_3^+$ toward the Galactic W51 by \cite{ind12} have indeed
  indicated a relatively low value of $\epsilon_{\mathrm{OH^+}}$.  
  In NGC~4418, however, PAH emission is very weak \citep{spo07}, and
  it is also weak in Arp~220 relative to the submillimeter continuum
  \citep{haa01}. Although this may be due to extinction of nuclear far-UV
  photons or foreground mid-IR extinction, rather than low PAH abundance, 
  we conservatively switch-off the charge exchange between gas
  and dust grains in our calculations. In this respect, the ionization rate
  $\zeta$ we derive below (Sect~\ref{sec:zeta}) should be viewed as a lower
  limit.

Under our simplified assumptions, and as shown by previous
studies \citep[e.g.][]{mal96,neu10,hol12}, the relevant input parameters are
$\zeta/n_{\mathrm{H}}$, $f_{\mathrm{H_2}}$, the gas temperature
($T_{\mathrm{gas}}$), and the oxygen abundance in the gas phase
($\chi_{\mathrm{O}}$). For the latter parameters we adopt
$T_{\mathrm{gas}}=150$ and $550$ K, and solar metallicities with 
$\chi_{\mathrm{O}}=3\times10^{-4}$. Freeze-out of atomic oxygen on 
  grains surfaces is thus ignored owing to the high \tdust\ in the nuclear
  regions of these sources \citep[e.g.][]{hol12}.

In the calculations, we treat $f_{\mathrm{H_2}}$ as an independent parameter,
even though its value can in principle be obtained from the H$_2$ formation
and destruction rates. Treating $f_{\mathrm{H_2}}$ as a free parameter has the
advantage that it allows us to run simple single-point models without additional
assumptions on the geometry or H$_2$ formation efficiencies. Furthermore,
$f_{\mathrm{H_2}}$ also depends on other physical processes that are difficult
to evaluate in the sources under study:
photodissociation by UV radiation, 
and the evolutionary state of the cloud in low density (diffuse)
clouds, where $f_{\mathrm{H_2}}$ only attains steady state after $\gtrsim10$
Myr \citep{lis07}. Concerning the 
latter point, the ground-state lines of \ohp\ and \hdop\ have been detected in
Galactic diffuse clouds indicating a very low $f_{\mathrm{H_2}}$
\citep{ger10,neu10}, probably reflecting regions that are evolving in
$f_{\mathrm{H_2}}$. Our calculations are intended to describe the
observations of both the nuclear regions and the extended component in Arp
220; in the latter case similar conditions to those found in Galactic
diffuse clouds may be present. For a given $\zeta/n_{\mathrm{H}}$, the
maximum value that $f_{\mathrm{H_2}}$ can attain is described
  as\footnote{It is assumed 
    that each H$_2$ ionization gives rise to one net H$_2$ destruction due to
    the expected dominant paths $\mathrm{H_2^++H_2\rightarrow H_3^++H}$ and
    $\mathrm{H_3^++e\rightarrow H_2+H}$.}:
\begin{equation}
f_{\mathrm{H_2}}^{\mathrm{max}}=\left(
1+\frac{\zeta/n_{\mathrm{H}}}{\gamma_{\mathrm H_2}} \right)^{-1}=
\left(1+\frac{\zeta/n_{\mathrm{H}}}{3.7\times10^{-17}\,
\mathrm{cm^3\,s^{-1}}} \right)^{-1},
\label{fh2max}
\end{equation}
where we have adopted for the rate coefficient of H$_2$ formation on
  grain surfaces the reference value $\gamma_{\mathrm H_2}=R_G T_g^{1/2}$ with
$R_G=3\times10^{-18}$ cm$^3$ s$^{-1}$ and $T_{\mathrm{gas}}=150$ K, and applied
this constant value to all cases (i.e. regardless of \tgas\ and \tdust)
for simplicity \citep[see discussion in][]{kau99}. 
For the $\zeta/n_{\mathrm{H}}$ values considered in the following
  sections, $5\times10^{-19}$, $5\times10^{-18}$, $2\times10^{-17}$,
$5\times10^{-17}$, and $5\times10^{-16}$ cm$^3$ s$^{-1}$,
$f_{\mathrm{H_2}}^{\mathrm{max}}$ is estimated as 0.99, 0.88, 
0.65, 0.43, and 0.07, respectively. Values of $f_{\mathrm{H_2}}$ below 
$f_{\mathrm{H_2}}^{\mathrm{max}}$ are obtained in our models by 
manually increasing the H$_2$ destruction rate. 
Equation~\ref{fh2max} appears to be roughly consistent with the models by
  \cite{bay11}, who obtain $f_{\mathrm{H_2}}\approx0.7$ at high $A_V$ for
  $\zeta/n_{\mathrm{H}}=10^{-17}$ cm$^3$ s$^{-1}$ and solar
  metallicities. However, the actual $f_{\mathrm{H_2}}^{\mathrm{max}}$ values 
  could be lower, owing to the decreasing H$_2$ recombination efficiency in
  warm grains \citep[$\gtrsim100$ K;][]{caz04,cup10}.
The grid of models cover
the $\zeta/n_{\mathrm{H}}-f_{\mathrm{H_2}}$ plane, with $f_{\mathrm{H_2}}$ varying
from 0.02 to $f_{\mathrm{H_2}}^{\mathrm{max}}$, and $\zeta/n_{\mathrm{H}}$
up to $5\times10^{-16}$ cm$^3$ s$^{-1}$.

   \begin{figure*}
   \centering
   \includegraphics[angle=-90,width=16.5cm]{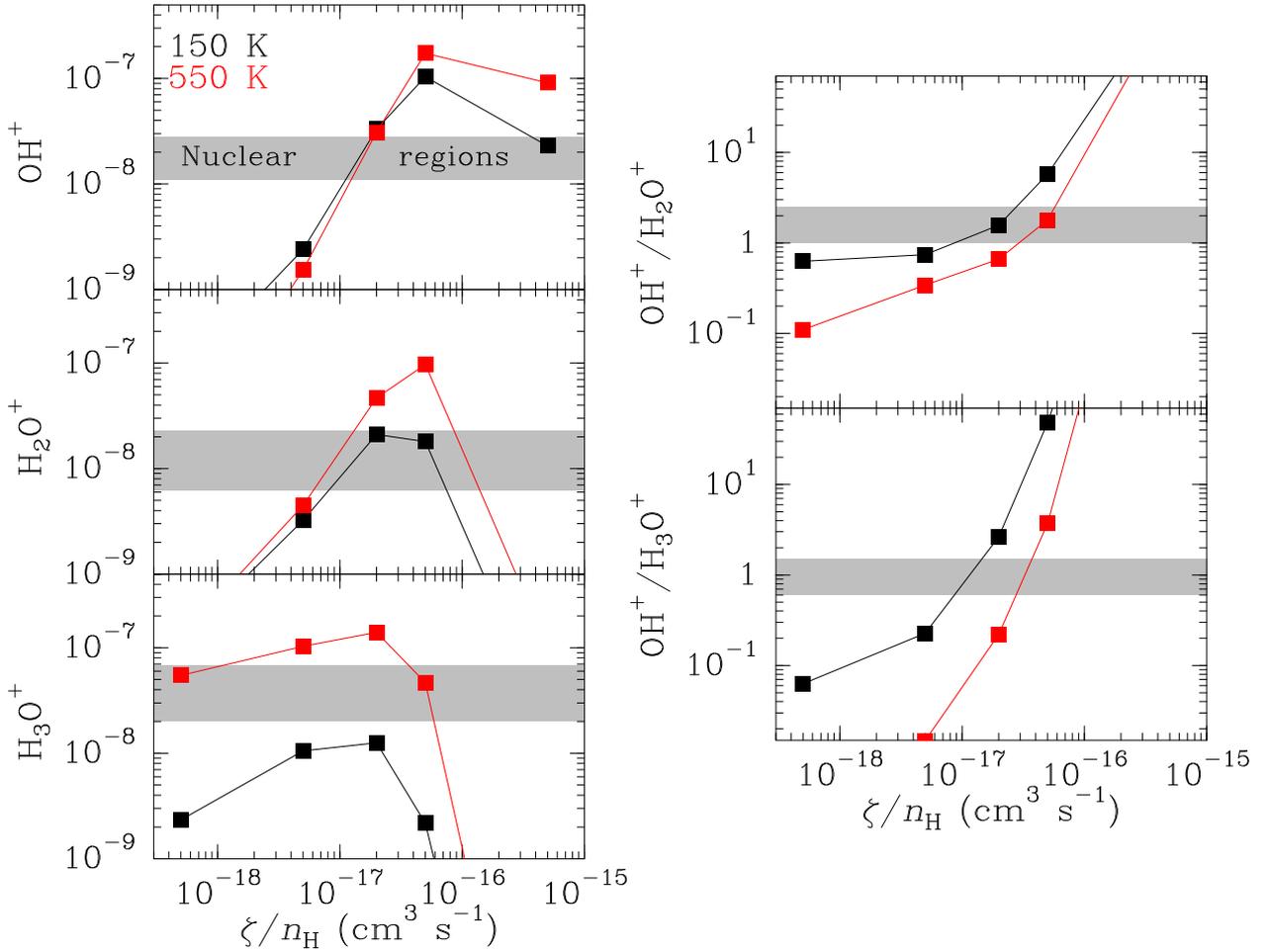}
   \caption{Predicted \ohp, \hdop, and \htop\ abundances (left panels)
     and \ohp/\hdop\ and \ohp/\htop\ abundance ratios (right panels) as
     a function of $\zeta/n_{\mathrm{H}}$. In all these models, the molecular
     fraction is given by the maximum value $f_{\mathrm{H_2}}^{\mathrm{max}}$ in
     eq.~(\ref{fh2max}). Gas temperatures of 150 K (black) and 550 K (red) 
     are assumed. The ranges derived from the observations and models for
     the nuclear regions of NGC~4418 and Arp~220 are indicated by the shaded
     regions.}
    \label{values_oh+_h2o+_h3o+}
    \end{figure*}

\subsection{The OH$^+$/H$_2$O$^+$ and OH$^+$/H$_3$O$^+$ ratios}

Figure \ref{ratio_oh+_h2o+} shows with squares, for all generated models, the
predicted $R_1\equiv\mathrm{OH^+}/\mathrm{H_2O^+}$ and
$R_2\equiv\mathrm{OH^+}/\mathrm{H_3O^+}$ ratios as a function of
$\chi_e\times10^3/f_{\mathrm{H_2}}$, where $\chi_e$ is the electron abundance
relative to H nuclei. As shown by \cite{ger10} and \cite{neu10}, if the
\hdop\ is exclusively formed by reaction of \ohp\ with H$_2$, $R_1$ only
depends on $\chi_e/f_{\mathrm{H_2}}$ and \tgas, regardless of the formation
mechanism of \ohp, according to:
\begin{equation}
R_1\equiv \mathrm{OH^+/H_2O^+} =0.63+0.85\times  
\frac{\chi_{\mathrm{e}}\times10^{3}}{\sqrt{T_{300}}f_{\mathrm{H_2}}},
\label{eq:e}
\end{equation}
where $T_{300}=T_{\mathrm{gas}}/300\,\mathrm{K}$. If in addition \htop\ is
exclusively formed from $\mathrm{H_2O^++H_2}$, $R_2$ is given by 
\begin{equation}
R_2\equiv \mathrm{OH^+/H_3O^+} =0.86\times 
\frac{\chi_{\mathrm{e}}\times10^{3}}{\sqrt{T_{300}}f_{\mathrm{H_2}}} \times
\left(1+1.34 \times 
\frac{\chi_{\mathrm{e}}\times10^{3}}{\sqrt{T_{300}}f_{\mathrm{H_2}}} \right).
\label{eq:e2}
\end{equation}
These analytical dependences are shown with solid curves for
$T_{\mathrm{gas}}=150$ and $550$ K in
Fig.~\ref{ratio_oh+_h2o+}. Eq.~(\ref{eq:e}) shows that a lower limit of 
$R=0.63$ is expected when recombination of \hdop\ is ignorable and the unique
route of \hdop\ destruction is reaction with H$_2$ yielding H$_3$O$^+$. For
$\chi_e/f_{\mathrm{H_2}}\gtrsim10^{-3}$, recombination of \hdop\
becomes increasingly important and $R_1$ rises linearly with
$\chi_e/f_{\mathrm{H_2}}$. This analytic approach to $R_1$ is appropiate for
regions where the formation of \hdop\ is dominated by 
$\mathrm{H_2 + OH^+}$. However, at low $\chi_e/f_{\mathrm{H_2}}$ values,
i.e. in shielded/high density regions with low electron fraction and
high molecular fraction, $R_1$ deviates from the analytic curve because there
are competing paths for \hdop\ formation (mostly $\mathrm{H^++H_2O}$ and
$\mathrm{H_3^++O}$) that do not involve formation of \ohp,
thus decreasing the $\mathrm{OH^+/H_2O^+}$ ratio below
$0.63$ (squares in Fig.~\ref{ratio_oh+_h2o+}a). On the other hand, $R_2$ varies
strongly with $\chi_e/f_{\mathrm{H_2}}$, and significant departures from the
analytical approach in eq.~(\ref{eq:e2}) are also obtained for low
$\chi_e/f_{\mathrm{H_2}}$, and high \tgas.

Even in the nuclear regions of NGC 4418 and Arp 220 where  densities are
high and high concentration of molecular species are observed
(Paper I), we do find relatively high values of $R_1$ and $R_2$,
though significantly lower than in Galactic diffuse clouds. $R_1$ is
found $\sim1-2$ and could be even higher in NGC 4418, and $R_2\sim1$
(Table~\ref{tab:columns}). Thus the inferred ratios appear to indicate that
the primary route for \hdop\ and \htop\ formation in the nuclear regions of
NGC 4418 and Arp 220 is the same as in Galactic translucent clouds,
$\mathrm{H_2 + OH^+}$ and $\mathrm{H_2 + H_2O^+}$, respectively, 
additional routes for \hdop\ and \htop\ formation being less important. 
We still expect, however, $f_{\mathrm{H_2}}$ to be significantly higher
than in translucent clouds. Adopting conservatively
$f_{\mathrm{H_2}}\gtrsim0.5$, we obtain
$\chi_{\mathrm{e}}\gtrsim2\times10^{-4}$. This value is  
higher than that expected from complete ionization of carbon,
$1.4\times10^{-4}$, pointing to a significant, additional source of free
electrons arising from H and H$_2$ ionization. 
In the more extended and diffuse \cext\ of Arp 220, the high value of $R_1$
probably reflects a relatively low value of $f_{\mathrm{H_2}}$, resembling the
situation found in Galactic translucent clouds.

The full chemical results for \ohp, \hdop, and \htop, together with the
corresponding abundances of electrons and H$^+$, are shown in
Appendix~\ref{appc}, where some analytical approaches to the abundance of
\ohp\ are also derived. We find that the most likely path for \ohp\ formation
is $\mathrm{O^++H_2}$, with the contribution from the charge exchange
reaction $\mathrm{H^++OH}$ and from direct OH ionization being of secondary
importance even in the nuclear regions, where high OH abundances are inferred
(Paper I). Therefore, the sequence 
$\mathrm{H^+ \rightarrow O^+\rightarrow OH^+\rightarrow H_2O^+\rightarrow
 H_3O^+}$ is favored as the production of the O-bearing cations. In
Appendix~\ref{appc} we also favor deep penetration of X-cosmic rays as
the main source of H$^+$ production, with the contribution due to
the $\mathrm{C^++OH\rightarrow CO^++H \rightarrow CO+H^+}$ FUV/chemical route
\citep{ste95,hol12} being presumably important only in small volumes around O
stars.

\subsection{The value of $\zeta/n_{\mathrm{H}}$ in the extended region of
  Arp~220}  

In the \cext\ component of Arp 220, where high \ohp\ column and $R_1$
values are found, our best fit values for $\zeta/n_{\mathrm{H}}$ 
and $f_{\mathrm{H_2}}$ are $(0.5-2)\times10^{-17}$ cm$^3$ s$^{-1}$ and
$0.05-0.25$, respectively (Fig.~\ref{chemres}). With this low molecular
fraction, however, $\chi_{\mathrm{OH^+}}$ is not far from saturation (see
Appendix~\ref{appc}), which occurs at $\chi_{\mathrm{OH^+}}\gtrsim10^{-7}$
and thus may be reached in some regions of \cext. The low
value of $f_{\mathrm{H_2}}$ indicates that either the H$_2$ 
abundance has not fully 
evolved to steady state, or H$_2$ photodissociation keeps a large hydrogen
fraction in atomic form. The electron abundance is expected 
to be in the range $\chi_{\mathrm{e}}\sim (3-10)\times10^{-4}$. 

These physical conditions are derived from the apparently high \ohp\
column in \cext, and from the relatively low \hdop\ column 
with the following caveat. The columns are derived from far-IR
\ohp\ absorption and submillimeter \hdop\ emission lines, 
which may probe different regions. The lowest-lying
absorption lines of \hdop\ at $183-184$ $\mu$m, observed in absorption toward
the nuclei (\chalo), indicate a foreground \hdop\ column of
$\sim3\times10^{15}$ \cmd. A fraction of the \htop\ observed
toward the nuclei is expected to arise also from foreground gas, with a column
of also $\mathrm{several}\times10^{15}$ \cmd, though detailed models for
  PDRs irradiated by cosmic rays show that \htop\ attains significant columns
  deeper into the cloud than \ohp\ \citep{hol12}.
The extended gas in Arp~220 likely harbors regions with a diversity of
physical conditions.

\subsection{The value of $\zeta/n_{\mathrm{H}}$ in the nuclear regions}

In the nuclear regions of NGC~4418 and Arp 220, where
$\chi_{\mathrm{OH^+}}\sim2\times10^{-8}$ and $R=1-2.5$
(Table~\ref{tab:columns}), $\chi_{\mathrm{OH^+}}$ is unsaturated
(see Appendix~\ref{appc}), i.e. results are sensitive to
$\zeta/n_{\mathrm{H}}$. However, the inferred $\zeta/n_{\mathrm{H}}$ depends
on the assumed $f_{\mathrm{H_2}}$. In these regions, we may expect steady
state for H$_2$, and we adopt $f_{\mathrm{H_2}}$ as given by
eq.~\ref{fh2max} (i.e. we assume that cosmic/X rays are the only source
  of gas dissociation/ionization). 
Figure~\ref{values_oh+_h2o+_h3o+} shows the corresponding 
modeled cation abundances and ratios. Model results can be summarized as
follows \citep[see also][]{neu10,ger10,gup10,hol12}. For very high
$\zeta/n_{\mathrm{H}}$, \ohp\ can still attain high 
abundances even with low $f_{\mathrm{H_2}}<10$\%, but \hdop\ is primarily
destroyed through recombination and the abundances of \hdop\ and \htop\ are
low. For low $\zeta/n_{\mathrm{H}}$, low abundances of \ohp\ and \hdop\ 
are obtained, but the \htop\ abundance for $T_{\mathrm{gas}}=550$ K 
remains high as \htop\ is formed through reaction of highly abundant
$\mathrm{H_2O}$ with $\mathrm{H_3^+}$ and $\mathrm{H^+}$.
Abundances of \ohp\ and \hdop\ around $10^{-8}$ are obtained at typical
$\zeta/n_{\mathrm{H}}\sim 10^{-17}$ cm$^3$ s$^{-1}$. The abundance of
\htop\ is very sensitive to \tgas, and $R_2$ is very sensitive to both
\tgas\ and $\zeta/n_{\mathrm{H}}$.

The inferred abundances and ratios, marked by shaded regions in
Fig.~\ref{values_oh+_h2o+_h3o+}, are  
consistent with $\zeta/n_{\mathrm{H}}$ in the 
range $(1-2)\times10^{-17}$ cm$^3$ s$^{-1}$. The electron abundance is
expected to be in the range $\chi_{\mathrm{e}}\sim (1.5-5)\times10^{-4}$.
These results are sensitive to the assumed H$_2$ fraction, with higher (lower)
$f_{\mathrm{H_2}}$ implying higher (lower) $\zeta/n_{\mathrm{H}}$. Decreasing 
$\gamma_{\mathrm H_2}$ in eq.~(\ref{fh2max}) to $10^{-17}$ cm$^3$ s$^{-1}$,
the observed abundances and ratios can be explained with  
$\zeta/n_{\mathrm{H}}\sim 5\times10^{-18}$ cm$^3$ s$^{-1}$. Further study of
the H$_2$ reformation on warm dust grains in these regions is required to
refine our estimates.

\subsection{The total ionization rate per H nucleus ($\zeta$)}
\label{sec:zeta}

Estimation of the total ionization rate per H nucleus $\zeta$ requires 
knowledge of the gas density. 
In the \cext\ of Arp 220, an average density of 
$n_{\mathrm{H}}\sim10^3$ cm$^{-3}$ is expected from the continuum emission and
estimated size (Paper I). Therefore $\zeta\sim10^{-14}$ s$^{-1}$, about 100
times higher than in Galactic diffuse clouds. This value is however 
uncertain, as the absorption in the \ohp\ lines may be selectively
produced in regions with densities much lower than the estimated average, and
thus $\zeta$ significantly lower than our above estimate is not ruled out.

In the nuclear regions of NGC 4418 and Arp 220, very high densities are 
expected. In the \ccore\ of NGC 4418, densities of $3\times10^6$ cm$^{-3}$ are
derived from the faint OH absorption in the ground-state lines at central
velocities, from the pattern of \hdo\ absorption, and from the high excitation
of HCN (Paper I). If the observed \ohp\ were formed in this region, this would
imply an extreme, very unlikely value of $\zeta\sim10^{-10}$ s$^{-1}$. 
However, it is plausible that the \ohp\ lines
are formed in a more tenous, nuclear region. There are several observations
indicating that the molecular ions are formed in a component different from
that responsible for the emission/absorption observed in neutral molecular
species. \cite{gup10} have detected broad
blueshifted absorption in the ground-state lines of \ohp\ and \hdop\
with HIFI toward the Orion KL, and their preliminary models appear to point
toward a component of the low-velocity Orion outflow that has a density
($\sim10^3$ cm$^{-3}$) far below the densities inferred from other species.
In diffuse gas, \cite{neu10} show that the velocity distribution of the
\ohp\ and \hdop\ absorption is dissimilar from that of \hdo; the
absorption in the molecular ions (and atomic H) being more broadly distributed
in velocity space. Theoretically, the models by \cite{bay11} and \cite{mei11}
indicate that the \ohp\ abundance is anticorrelated with the abundance of
NH$_3$ and HCN. Furthermore, the models by \cite{mei11} indicate that the
\hdo\ abundance is anticorrelated with $\zeta/n_{\mathrm{H}}$, and thus with
the abundances of \ohp\ and \hdop\ as well.

We thus conservatively propose
that the absorption observed in the high-lying lines of \ohp\ and \hdop\ are
formed in a relatively low-density region different from the dense component
that accounts for the absorption in the high-lying lines of \hdo, HCN, and
NH$_3$ (Paper I). This low density component could be interpreted in terms of an
nuclear interclump medium, or a shell surrounding the high density nuclear
gas. However, this component cannot be very extended as
otherwise the far-IR radiation density could not excite the molecular ions up
to the observed rotational levels ($E_{\mathrm{lower}}\approx285$ K). 
The width of the \ohp\ shell should be small in comparison with the radius of
the far-IR source, which provides a lower limit on the density.  
Our models favor densities of at least $\sim10^4$
\cmt\ in Arp~220, and $\sim2\times10^4$ \cmt\ in NGC~4418. For a
density threshold of $10^4$ cm$^{-3}$, $\zeta\gtrsim(1-2)\times10^{-13}$
s$^{-1}$. In NGC~4418, where high densities were inferred in Paper I, 
this is a strong lower limit, and yet still
$\zeta\gtrsim\mathrm{several}\times10^2$ times the
highest values estimated in the Milky Way
\citep[$\mathrm{several} \times10^{-16}$ s$^{-1}$;
  e.g.][]{ind07,neu10,hol12,ind12}.   

\section{Discussion}
\label{sec:discussion}

\subsection{The O-bearing cations and the neutral species}
\label{neutral}

Under the simple assumptions used in our schematic chemical models,
predictions for the abundances of the neutral species OH, \hdo, \nht, and HCN
analyzed in Paper I are also obtained. 
The main results for the neutral species are: $(i)$ OH attains abundances
$>10^{-6}$ even for the lowest $\zeta/n_{\mathrm{H}}=5\times10^{-19}$ cm$^3$
s$^{-1}$ and a wide range in $f_{\mathrm{H_2}}$, in rough agreement with the
derived values.  
$(ii)$ \hdo\ only attains the inferred high abundances of $\sim10^{-5}$ 
for high \tgas\ and $\zeta/n_{\mathrm{H}}<2\times10^{-17}$ cm$^3$ s$^{-1}$;
otherwise gas-grain chemistry (not included in our simple chemical
calculations) is required to account for the high \hdo\ abundance (Paper I). 
$(iii)$ HCN only approaches abundances as high as $\sim10^{-6}$ for
the lowest $\zeta/n_{\mathrm{H}}=5\times10^{-19}$ cm$^3$ s$^{-1}$ and high
\tgas, suggesting that indeed it is formed in regions with densities higher,
and probably more protected from the ionization source, than those responsible
for the molecular ions. $(iv)$ \nht\ attains abundances of only $10^{-8}$, far
below the inferred ones, further indicating the importance of gas-grain
chemistry. In general, clumpy regions with high contrast in $n_{\mathrm{H}}$,
$f_{\mathrm{H_2}}$, and $\zeta$, and gas-grain chemistry, are most likely
required to account for the full molecular absorption detected in the nuclear
regions of both galaxies. The neutral/ion connection will be further analyzed
in a future study.

\subsection{X-rays or cosmic rays?}
\label{sec:xc}

In the starburst galaxies NGC~253 and M82, 
detection of $\gamma-$rays \citep{ace09,acc09,abd10} indicate a
high CR energy density of $\sim200-300$ times the value ($\sim1$ eV/cm$^3$) in
the Milky Way \citep{abr12,per12}.
Since the $\gamma-$rays are thought to be the result of cosmic rays
accelerated by supernova remnant shocks that interact with high density
target material, a high rate of cosmic-ray ionization indicates a high
rate of supernovae in the central region of starburst galaxies 
  \citep[$\nu_{\mathrm{SN}}\sim0.03$ yr$^{-1}$ in NGC~253,][]{eng98}, which
would be expected to be accompanied by observable galactic outflows. These are
indeed seen in both NGC~253 and M82 \citep[e.g.][]{sug03,str07,vei09}.  

The cosmic-ray energy density inferred in Arp~220 from both the supernova
  rate \citep[$\sim4$ yr$^{-1}$,][]{lon06} and
synchrotron radio emission is $\sim500-1000$ times the Galactic
value \citep{per12}, in good agreement with our lower limit of 
$\mathrm{several\times10^2}$ times the Galactic value for $\zeta$.
The blueshifted component in Arp~220 traced by
the high-excitation lines of \hdo, OH, \nht,
and HCN may indicate outflowing gas with a line of sight velocity of
$\sim75$ \kms\ (Sect.~\ref{sec:vel_arp220}). 
With an extent of $\sim80$ pc, this would 
imply a ({\em local}) very high mass outflow rate of
$\sim900\times(n_{\mathrm{H}}/10^5\,\mathrm{cm^{-3}})$ 
\Msun\ yr$^{-1}$. However, the outflowing gas is difficult to
distinguish from the rotating gas for the relatively low outflow 
velocities in Arp~220 (Sect.~\ref{sec:vel_arp220}), 
in stark contrast with the $\sim1000$ \kms\ outflow velocities traced by
OH in other ULIRGs \citep{fis10,stu11}. Nevertheless, clear 
indication of outflowing gas is seen in the P-Cygni HCO$^+$ profiles
observed at submillimeter wavelengths \citep{sak09}, and is also suggested
at redshifted velocities in the [O {\sc i}] 145 $\mu$m line and in the
low-lying OH and \hdo\ emission component profiles (Paper I). 

The above independent estimates of the CR energy densities in Arp~220,
  and the signatures of outflowing molecular gas with low-moderate velocities,
appear to be compatible with the 
scenario of a high density of cosmic rays producing the observed molecular
ions. \cite{pap10} has proposed that very high cosmic ray energy densities, 
of the order of those we derive here, can turn the bulk of the nuclear
molecular gas in ULIRGs into giant CR-dominated Regions (CRDRs) rather than
ensembles of Photon-dominated Regions (PDRs). In this framework, the O-bearing
cations can be considered excellent tracers of these regions. 

The case of NGC~4418 is unclear. 
This highly buried \citep{spo01,spo07} and highly excited (Paper I) source,
where the bulk of the luminosity arises from a very small
volume, is deficient in synchrotron emission with respect to the far-infrared
emission \citep{rou03}.  
Furthermore, we did not find spectroscopic signatures of outflowing gas; on
the contrary the far-IR lines observed with PACS are narrow (Paper I).
Therefore, cosmic rays may only be of secondary importance.
An explanation of the synchrotron deficit based on a nascent starburst
\citep{rou03} can hardly account for the excited \ohp\ detected in the source,
though observations of excited \ohp\ in Galactic and other extragalactic
sources are needed to check this point. Preliminarily,
NGC~4418 is better interpreted as a buried AGN, and the molecular 
ions as produced in a X-ray Dominated Region (XDR). But given that Arp~220 can
be considered as a scaled-up version of NGC~4418, and that the columns of the
cations are found similar in the nuclear regions of both sources, the possible
XDR nature of Arp~220 is not ruled out.

While both NGC~4418 and Arp~220
  are underluminous in hard ($2-10$ keV) X-rays for their computed SFRs
  \citep{mai03,iwa05}, the X-ray {\em absolute} fluxes may suffice to
  produce the inferred ionization rate. The distance-corrected X-ray fluxes of
  $0.1$ and $0.5$ erg s$^{-1}$ cm$^{-2}$ estimated at the surface of the
  nuclear regions in NGC~4418 and Arp~220 (Paper I),
  respectively, are lower limits due to foreground absorption in both galaxies
  (with $N_{\mathrm{H}}\sim10^{23}$ cm$^{-2}$, Paper I) and by the molecular shell
  itself where the ions reside. Depending on the X-ray spectrum, {\em local}
  (i.e. unattenuated) X-ray fluxes of $\gtrsim0.5$ erg s$^{-1}$ cm$^{-2}$ may
  well produce $\zeta$ in excess of $10^{-13}$ s$^{-1}$.

The strong high-ionization Fe K line in Arp~220 \citep{iwa05} appears to
indicate that the faint X-ray emission is not primarily due to high-mass X-ray
binaries (as in starburst galaxies), but could 
  arise either from an internally shocked hot bubble of starburst origin (the
  pressure of which would produce the observed molecular outflow), or from a
  deeply embedded supermassive black hole \citep{iwa09}. 
  In the latter case, the molecular ions would then be formed in (somewhat
  extended) regions that are illuminated by X-rays from the AGN that can
  escape to the outer (nuclear) regions primarily through inhomogeneities in
  the nuclear media. The drawback of this alternative is that the far-IR
  molecular lines are observed in absorption against dust emission that is
  optically thick even at far-IR wavelengths, and thus the X-ray emission from
  a putative buried AGN would then be strongly attenuated in arriving at such
  nuclear, but ``surface'' regions unless some special geometry is
  invoked\footnote{The 
    extinction in the X-ray regime for $E_X<10$ keV is higher than at 200
    $\mu$m for a gas-to-dust ratio of 100 by mass \citep[e.g.][]{mon01}.}.
 Further theoretical and observational studies are required to 
   discriminate between CRDRs and XDRs.

\subsection{\htop: formation pumping or hot gas?}

We show in Appendix~\ref{appb} that the \htop\ population diagram
for the PIMS lines in Arp~220 can be understood in terms of formation pumping
described by $T_{\mathrm{form}}\sim1000$ K, but that plausible 
collisional relaxation and the lack of knowledge of collisional rates make
the case uncertain. Interestingly, detailed chemical models of PDRs
\citep{hol12} predict that \htop\ is primarily formed in fully molecular
regions, while both \ohp\ and \hdop\ peak closer to the PDR surface where 
$f_{\mathrm{H_2}}$ is significantly lower than unity. On the contrary, in
Arp~220, the \htop\ PIMS lines peak at the redshifted velocity (LE) component
(Fig.~\ref{vel}), which is expected to be more widespread and less dense than
the blueshifted velocity (HE) component. 

Nevertheless, it is worth noting that \cite{ran11} have
detected high-$J$ rotational lines of CO, which the authors interpreted as
arising from hot ($\sim10^3$ K), low density ($\sim10^3$ cm$^{-3}$) gas that is
mechanically heated. If the \htop\ lines are formed in the same region as the
CO lines, the presence of such a hot molecular gas layer in Arp~220 may indeed
suggest that {\em both} formation pumping {\em and} high \tgas\ are
responsible for the observed PIMS absorption. 
Due to the low gas density associated with the redshifted (LE) component,
  inefficient cooling may keep the gas at high \tgas, which has the additional
  effect of boosting the \htop\ abundance (Fig~\ref{values_oh+_h2o+_h3o+}). On
  the other hand, the nuclear gas associated with the blueshifted velocity (HE)
  component is dense, and gas-grain thermal coupling \citep{gol01} 
  may be efficient in cooling the gas to the inferred $150-200$ K
  (based on \nht\ lines, Fig.~\ref{nh3h2o}b). Collisional relaxation of
  the \htop\ metastable populations may proceed efficiently at this \tgas\
  (Fig.~\ref{h3o+form}). 

For a density of $10^4$ cm$^{-3}$ and $\zeta=10^{-13}$ s$^{-1}$, the expected
\tgas\ is $\sim300$ K and rises very steeply with increasing $\zeta$
\citep{bay11}; this result suggests X/cosmic ray heating may be important, 
together with mechanical heating, in explaining \tgas\ in the nuclear regions
of NGC~4418 and Arp~220.

\section{Conclusions}
\label{sec:conclusions}

The observational results on \ohp, \hdop, and \htop\ in NGC~4418 and Arp~220
  are: 
\begin{itemize}
\item Absorption lines from excited levels of OH$^+$ with $E_{\mathrm{lower}}$
  up to 285 K are detected in NGC~4418 and Arp~220. While the highest-lying
  lines have similar strengths in both sources, the low-lying lines are more
  prominent in Arp~220. 
\item Lines of \hdop\ are also well detected in absorption in Arp~220 up to a
  lower level energy of $\sim200$ K, with strong absorption in the
  ground-state p-\hdop\ transition at $183-184$ $\mu$m. 
  In contrast, in NGC~4418 the ground-state line is not detected and only
    several of the excited lines are detected with relatively low
    signal-to-noise.
\item The pure-inversion, metastable lines of \htop\ are detected in Arp~220
  up to a lower level of $\sim1400$ K, indicating a rotational temperature of
  $\sim550$ K. These lines are not detected in NGC~4418. However, the
  rotation-inversion lines, some of them arising from non-metastable levels,
  are detected in both sources. 
\item We identify two main velocity components in Arp~220 separated by
  $\approx75$ \kms. The high-lying lines of all neutral (\hdo, OH, \nht, HCN)
  and ionized (\ohp, \hdop) molecular species peak at the blueshifted
  component, while the lowest-lying lines peak at the redshifted one. The
  exception corresponds to the pure-inversion metastable lines of \htop, which
  also peak at the velocity of the redshifted component. The blueshifted line
  components may be tracing the outflow shown in the millimeter lines of
    HCO$^+$ by \cite{sak09}, and grain mantle evaporation, sputtering, and
    neutral-neutral reactions at high \tgas\ may all be contributing to
    enhance the gas-phase abundances of \hdo\ and \nht.
\end{itemize} 

The observations are analyzed with radiative transfer modeling. The
conclusions stemming from the analysis are: 
\begin{itemize}
\item The excited lines of \ohp\ and \hdop, and the rotation-inversion lines
  of \htop\ are pumped through absorption of far-IR radiation emitted by dust,
  but the pure-inversion metastable lines of \htop\ are either
  chemically and/or collisionally pumped.
\item OH$^+$ column densities of $(0.5-1.1)\times10^{16}$ \cmd\ are derived
  toward the nuclear region of both sources for spatial extents of $20-30$ pc in
  NGC~4418, and $90-160$ pc in Arp~220. The estimated abundance is
  $(1-3)\times10^{-8}$ in both sources. In addition, in Arp~220, absorption
  from an extended region is required to match the low-lying \ohp\ lines, with
  an estimated column of $\mathrm{a\,few\times10^{16}}$ \cmd\ and an
  abundance of $\mathrm{several\times10^{-8}}$. 
\item The nuclear columns of \hdop\ are $(0.3-0.9)\times10^{16}$ \cmd\ in
  Arp~220, and $(0.3-0.4)\times10^{16}$ \cmd\ in NGC~4418, indicating
  $\mathrm{OH^+/H_2O^+}$ ratios of $1-2.5$ (the lower value applies to
  Arp~220). In Arp~220, the emission lines detected with SPIRE indicate
  an additional extended region with columns of $\mathrm{several\times10^{15}}$,
  similar to the column derived from the ground-state p-\hdop\ lines,
  indicating a $\mathrm{OH^+/H_2O^+}$ ratio of $5-10$ in the extended
    region. 
\item The \htop\ lines indicate columns of $(0.5-0.8)\times10^{16}$ \cmd\ in
  NGC~4418, and $(1-2)\times10^{16}$ \cmd\ in Arp~220. In Arp~220, the
  observed high excitation of the pure-inversion metastable lines may be due
  to both formation pumping and high \tgas, and the \htop\ lines are likely
  tracing a nuclear transition region into the more extended, low excitation
  gas. It is suggested that collisional relaxation of the \htop\ metastable
  populations is important in the dense nuclear gas of Arp~220. The \htop\
  lines are redshifted relative to the high-lying lines of \ohp\ with
  uncertain velocity overlap, and we estimate a $\mathrm{OH^+/H_3O^+}$ ratio
  of $\sim1$ in the nuclear region of both sources. 
\end{itemize}

Simple chemical models are used to interpret the results of radiative transfer
modeling. Our main conclusions are:
\begin{itemize}
\item The sequence $\mathrm{H^+ \rightarrow O^+\rightarrow OH^+\rightarrow
    H_2O^+\rightarrow H_3O^+}$ is favored as the primary route to explain the
  high columns of the O-bearing molecular ions toward the nuclear region of
  both sources.
\item The production of H$^+$ is most probably dominated by X/cosmic ray
  ionization, rather than by FUV/chemical means (Appendix~\ref{appc}). 
  An ionization rate per hydrogen nuclei density of
  $\zeta/n_{\mathrm{H}}\sim(1-2)\times10^{-17}$ cm$^3$ s$^{-1}$ is estimated
  in the nuclei of both galaxies. 
\item The full set of observations and models lead us to propose that the
    molecular ions arise in a relatively low-density interclump medium
   but with density of at least $10^4$ \cmt, in which case
   the ionization rate is $\zeta>10^{-13}$ s$^{-1}$, a
   lower limit that is over two orders of magnitude greater than the highest
     rate estimates of $\mathrm{several} \times10^{-16}$ s$^{-1}$ for Galactic
     regions.   
\item The cosmic-ray energy density inferred in Arp~220 from both the
    supernova rate and synchrotron radio emission is $\sim10^3$ times the
    Galactic value \citep{per12}, which is compatible with the lower
    limit found for $\zeta$ from the present observations and
    analysis. Nevertheless, an important contribution due to X-ray ionization
    in Arp~220 is also plausible. In NGC~4418, an X-ray Dominated Region is 
    favored. Further theoretical and observational studies are required to 
    discriminate between X-ray and cosmic ray ionization.
\end{itemize}

\begin{acknowledgements}
We thank the referee, David A. Neufeld, for useful indications that improved
the manuscript. PACS has been developed by a consortium of institutes
led by MPE (Germany) and including UVIE (Austria); KU Leuven, CSL, IMEC
(Belgium); CEA, LAM (France); MPIA (Germany); 
INAFIFSI/OAA/OAP/OAT, LENS, SISSA (Italy); IAC (Spain). This development
has been supported by the funding agencies BMVIT (Austria), ESA-PRODEX
(Belgium), CEA/CNES (France), DLR (Germany), ASI/INAF (Italy), and
CICYT/MCYT (Spain). E.G-A is a Research Associate at the Harvard-Smithsonian
Center for Astrophysics, and thanks the support by the Spanish 
Ministerio de Ciencia e Innovaci\'on under project AYA2010-21697-C05-01. Basic
research in IR astronomy at NRL is funded by 
the US ONR; J.F. also acknowledges support from the NHSC. 
H.S.P.M. acknowledges support by the 
Bundesministerium f\"ur Bildung und Forschung (BMBF) through project 
FKZ 50OF0901 (ICC HIFI \textit{Herschel}).
S.V. thanks NASA for partial support of this research via Research Support
Agreement RSA 1427277, support from a Senior NPP Award from NASA, and
his host institution, the Goddard Space Flight Center.
This research has made use of NASA's Astrophysics Data System (ADS)
and of GILDAS software (http://www.iram.fr/IRAMFR/GILDAS).
\end{acknowledgements}

\begin{appendix}
\section{Excitation of \ohp\ and \hdop\ through formation pumping}
\label{appa}

Here we show that formation pumping is not important for the \ohp\ and
  \hdop\ level population distribution or absorption line fluxes in Arp~220
  and NGC~4418. The effect of formation pumping on the excitation of \ohp\
can be estimated from the statistical equilibrium equations by \cite{bru10b}
(eqs. B1-B3 in their Appendix B), by ignoring the collisional and 
induced emission terms:
\begin{eqnarray}
\frac{dn_{N,J}}{dt} & = & \sum_{J'} n_{N+1,J'} A_{N+1,J';N,J} - 
n_{N,J} \sum_{J'} A_{N,J;N-1,J'} \nonumber \\
& + & \Gamma_{N,J} - n_{N,J} \Lambda_{N,J} = 0,
\label{eq:stateq}
\end{eqnarray}
where $A_{N,J;N',J'}$ is the Einstein coefficient for spontaneous emission
from level $(N,J)$ to level $(N',J')$, and $\Gamma_{N,J}$ (cm$^{-3}$ s$^{-1}$)
and $\Lambda_{N,J}$ (s$^{-1}$) are the chemical formation and destruction rates. 
The total \ohp\ formation rate is
\begin{equation}
\Gamma_{\mathrm{OH^+}}  =  \sum_{N,J} \Gamma_{N,J} 
  =  n_{\mathrm{H}} \epsilon_{\mathrm{OH^+}} \zeta,
\label{eq:formohplus}
\end{equation}
where $\zeta$ is the ionization rate and $\epsilon_{\mathrm{OH^+}}$ is the
efficiency with which ionizations are transfered to \ohp\ production. The
destruction rate $\Lambda_{N,J}$ is assumed to be independent of the considered
level \citep{bru10b}, and for \ohp\ is given by 
\begin{equation}
\Lambda_{N,J} = \Lambda_{\mathrm{OH^+}} = n_{\mathrm{H}} (\chi_{\mathrm{e}}
k_{\mathrm{OH^+|e}} +  0.5 f_{\mathrm{H_2}} k_{\mathrm{OH^+|H_2}}).
\label{eq:destohplus}
\end{equation}

In steady state, 
$\Lambda_{\mathrm{OH^+}}=\Gamma_{\mathrm{OH^+}}/(n_{\mathrm{H}} \chi_{\mathrm{OH^+}})$,
which is expected to be $<10^{-5}$ s$^{-1}$. Since this value is
much lower than the $A_{N,J;N',J'}$ values ($\gtrsim0.05$ s$^{-1}$,
Table~\ref{tab:fluxes}), the \ohp\ molecules are
destroyed once they are in the ground $(N,J)=(0,1)$ level,
and eq.~(\ref{eq:stateq}) yields for excited $N>0$ levels 
\begin{equation}
n_{N,J} \sum_{J'} A_{N,J;N-1,J'} = \Gamma_{N,J} + \sum_{J'} n_{N+1,J'}
A_{N+1,J';N,J}.
\label{eq:pobnj}
\end{equation}
For the ground $(N,J)=(0,1)$ level we can write: 
\begin{equation}
n_{0,1} \Lambda_{\mathrm{OH^+}} = \Gamma_{0,1} + \sum_{J'} n_{1,J'} A_{1,J';0,J}.
\label{eq:pob01}
\end{equation}
Equations~(\ref{eq:pobnj}) and (\ref{eq:pob01}) are easily solved iteratively
by starting from a high $N$-level down to the $N=0$ state. 
Results show that for $\chi_{\mathrm{OH^+}}\gtrsim10^{-8}$, $\zeta<10^{-12}$
s$^{-1}$, and $\epsilon_{\mathrm{OH^+}}\sim0.5$, the three $N=1\rightarrow0$
components have an excitation temperature $T_{ex}<7$ K, implying that less than
$0.3$\% of \ohp\ molecules are in excited levels. Therefore, excitation by
dust emission is much more efficient than formation pumping in these sources.
A similar analysis was done for \hdop, with also similar results:
$T_{ex}\lesssim7$ K for the ground state transitions, and less than $1.5$\% of
\hdop\ molecules in excited levels. 

The flux predicted in the \ohp\ $N\rightarrow N-1$ transition as derived from
formation pumping alone can also be easily estimated.
By summing over all $J-$levels in eq.~(\ref{eq:pobnj}), one obtains
\begin{equation}
\sum_{J,J'} n_{N,J} A_{N,J;N-1,J'}= \sum_{N'\ge N} \sum_{J} \Gamma_{N',J},
\end{equation}
thus the number of $N\rightarrow N-1$ line photons emitted per unit volume and
time is equal in steady state to the accumulated formation rate in all
$N'\ge N$ levels, and is independent of the $A_{ul}-$values. Writing the
volume of the emitting region as $V=\pi R^2 N_{\mathrm{H}}/n_{\mathrm{H}}$,
where $N_{\mathrm{H}}$ and $n_{\mathrm{H}}$ 
are the column and density of H nuclei, respectively, and 
$R$ is the effective radius of the region, the flux in the $N\rightarrow N-1$
fine-structure lines due to formation pumping is given by
\begin{equation}
F_{N\rightarrow N-1}= 
\frac{h \nu_{N\rightarrow N-1} R^2 N_{\mathrm{H}}}{4 n_{\mathrm{H}} D^2 }
\times \sum_{N'\ge N} \sum_{J} \Gamma_{N',J},
\label{eq:fluxn}
\end{equation}
where $D$ is the distance to the source. The double sum in eq.~(\ref{eq:fluxn})
is evidently lower than the total \ohp\ formation rate
($\Gamma_{\mathrm{OH^+}}$), and then 
\begin{equation}
F_{N\rightarrow N-1}< 
\frac{h \nu_{N\rightarrow N-1} R^2 N_{\mathrm{H}}\Gamma_{\mathrm{OH^+}}}
{4 n_{\mathrm{H}} D^2 },
\label{eq:fluxn2}
\end{equation}
and using eq.~(\ref{eq:formohplus}),
\begin{eqnarray}
F_{N\rightarrow N-1} & < & 
\frac{h \nu_{N\rightarrow N-1} R^2 N_{\mathrm{H}} \epsilon_{\mathrm{OH^+}}
  \zeta}{4  D^2 } \nonumber \\
& \sim &
1.3\times10^{-23} \epsilon_{\mathrm{OH^+}} 
\left(\frac{150\,\mathrm{\mu m}}{\lambda} \right)
\left(\frac{R}{50\,\mathrm{pc}} \right)^2 
\left(\frac{72\,\mathrm{Mpc}}{D} \right)^2 \nonumber \\
& \times &
\left(\frac{N_{\mathrm{H}}}{4\times10^{23}\,\mathrm{cm^{-2}}} \right)
\left(\frac{\zeta}{2\times10^{-13}\,\mathrm{s^{-1}}} \right)
\,\mathrm{W\,cm^{-2}}.
\label{eq:fluxn3}
\end{eqnarray}
The reference value is $\sim500$ times lower than the absorbing flux of the
$N=2\leftarrow1$ transition in Arp~220 (Table~\ref{tab:fluxes}), indicating that
formation pumping has a negligible effect on the observed line fluxes.
For NGC~4418, the upper limit is $3\times10^{-23}\,\mathrm{W\,cm^{-2}}$ for
$R=20$ pc, also much lower than observed. 

\end{appendix}

\begin{appendix}
\section{Excitation of \htop\ through formation pumping}
\label{appb}

   \begin{figure}
   \centering
   \includegraphics[width=8.5cm]{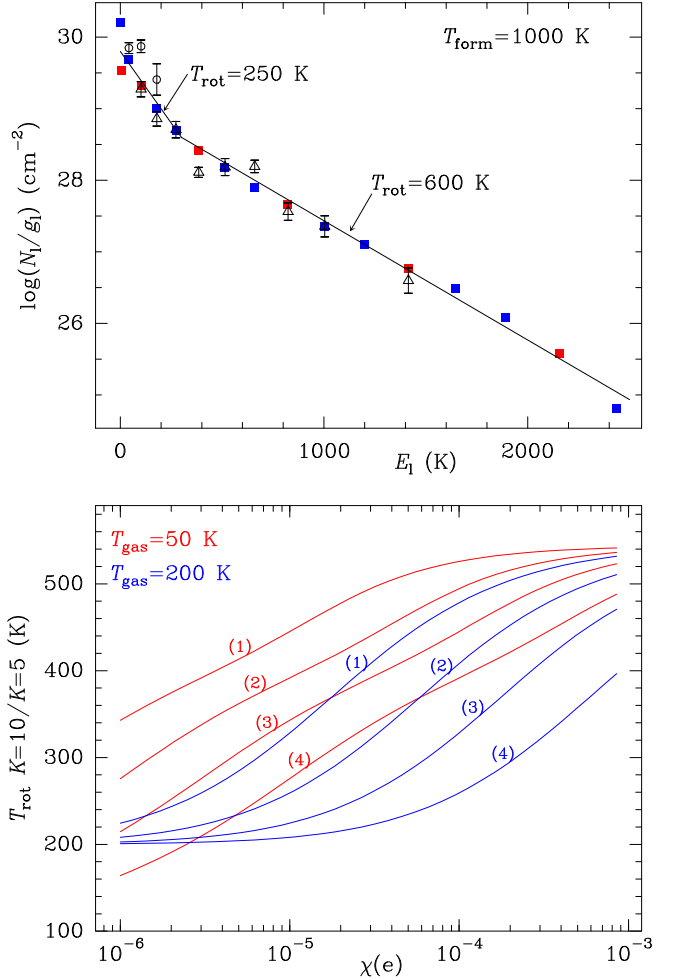}
   \caption{(a) The \htop\ rotation diagram (triangles), calculated in the
     optically thin limit. The colored squares (red: ortho, blue: para) show
     the expected column density distribution for a formation temperature of
     $T_{\mathrm{form}}=1000$ K, and neglecting collisional relaxation. The
     two solid lines correspond to rotational temperatures of
     $T_{\mathrm{rot}}=250$ and $600$ K. (b) The rotational temperature of the
   $10_{10}^+$ metastable level of \htop\ relative to the $5_5^+$ one as a
   function of the electron abundance $\chi(\mathrm{e})$ for two adopted
   values of $T_{\mathrm{gas}}$ (50 and 200 K) and four assumed values of
   $k_{K,K'}^{\mathrm{col}}=3\times10^{-11}$ (curves 1), $1\times10^{-10}$ (2),
   $3\times10^{-10}$ (3), and $1\times10^{-9}$ cm$^3$ s$^{-1}$ (4).}    
    \label{h3o+form}
    \end{figure}

As discussed in Sect.~\ref{sec:excit}, the rates for \htop\ excitation through
collisions are unknown and therefore the excitation mechanism of the
high metastable levels cannot currently be determined. Nevertheless,
here we attempt to constrain under what conditions formation
pumping can account for the observed excitation. 
By assuming a range of reasonable collisional excitation rates and some
  simplifying but illustrative assumptions, we show here that formation pumping
  can play an important role in populating the highly excited metastable
  levels from which absorption is observed in Arp~220.

We assume that the formation
distribution can be described by a formation temperature, $T_{\mathrm{form}}$,
so that the formation rate $\Gamma_i$ in level $i$ is \citep{bru10b}: 
\begin{equation}
\Gamma_{i}  = \Gamma_{\mathrm{H_3O^+}} \times \frac{g_i
  \exp\{-E_i/T_{\mathrm{form}}\}}{Q(T_{\mathrm{form}})},   
\label{eq:formoh3oplus}
\end{equation}
where $\Gamma_{\mathrm{H_3O^+}}$ is the total \htop\ formation rate, $g_i$ and
$E_i$ are the degeneracy and energy of level $i$, respectively, and $Q$ is the
partition function. \htop\ is expected to be formed through
$\mathrm{H_2O^++H_2\rightarrow H_3O^++H}$, with an exothermicity of $1.7$ eV.
Imposing equal momentum of the products in the rest frame, and assuming energy
equipartition of \htop\ between rotational, vibrational, and translational
energy, we roughly estimate that $\sim0.08$ eV per formed \htop\ molecule will
go into rotational energy, implying $T_{\mathrm{form}}\sim800$ K. We adopt in
the following $T_{\mathrm{form}}=1000$ K. 

After a molecule is formed in a
$J_K^p$ level, it will cascade down very quickly to the metastable $K_K^+$
one, where it will remain until a collisional event induces a
transition to another level, 
or until the molecule is destroyed through recombination with
an electron. If the latter process dominates, the population in the metastable
levels will be determined by the formation process. However, the rotational
temperature \trot\ of the metastable levels will be lower than \tform, because
the former is determined by the {\em J-accumulated} formation rates in the
different $K-$ladders. In Fig.~\ref{h3o+form}a, the colored symbols show the
expected population diagram obtained by ignoring collisional relaxation. While
there is not a unique \trot, there is a tendency for it to increase with
$E_l$, and the expected distribution can be roughly fitted with
$T_{\mathrm{rot}}\sim250$ and $\sim600$ K for $E_l\lesssim300$ K and 
$E_l>300$ K, respectively. This resembles the double distribution found by
\cite{lis12} toward Sgr~B2. The triangles in Fig.~\ref{h3o+form}a show the
columns obtained toward Arp~220 by assuming that the observed lines are
optically thin (Sect.~\ref{obs:h3o+}).

In order to check under which conditions the formation process can generate
the observed high excitation of the metastable \htop\ levels in environments
with moderate \tgas, we have made calculations for the equilibrium populations
$n_{K}$ of the metastable levels by solving the rate equations 
\begin{eqnarray}
\frac{dn_{K}}{dt} & = & \Gamma_{K}+ n_{\mathrm{H}} \sum_{K'} n_{K'}
k_{K',K}^{\mathrm{col}} 
\nonumber \\ 
& - & n_{K} \, n_{\mathrm{H}}\, 
\left( \sum_{K'} k_{K,K'}^{\mathrm{col}} + \chi_{\mathrm{e}}
  k_{\mathrm{H_3O^+|e}} \right) = 0,  
\label{eq:stateqh3op}
\end{eqnarray}
where $\Gamma_{K}$ denotes the accumulated formation rate in all levels of
the $K-$ladder. The rate coefficient for \htop\ destruction through
recombination is taken from the UMIST06 database \citep{woo07}:  
\begin{equation}
k_{\mathrm{H_3O^+|e}}=4.32 \times 10^{-7} \times 
\left( \frac{T_{\mathrm{gas}}}{300 \,\mathrm{K}} \right)^{-1/2} \, 
\mathrm{cm^3 \, s^{-1}}.
\label{eq:htoprec}
\end{equation}
We treat the collisional relaxation in a rather simplistic, but still
illustrative way: only collisions between a given metastable level and the
neighboring upper and lower (ortho or para) metastable levels are considered,
with an effective rate of collisional de-excitation from level $K$ to level
$K'$ given by $n_{\mathrm{H}}\, k_{K,K'}^{\mathrm{col}}$, assumed independent
of $K$. The excitation rate is fixed by the requirement of detailed balance at
the adopted \tgas.

We solved eqs.~(\ref{eq:stateqh3op}) for all metastable levels by
assuming an \htop\ ortho-to-para abundance ratio of 1, and for four
adopted values of $k_{K,K'}^{\mathrm{col}}$: $3\times10^{-11}$,
$1\times10^{-10}$, $3\times10^{-10}$, and $1\times10^{-9}$ cm$^3$ s$^{-1}$.
In steady state, $\Gamma_{K}\propto n_{\mathrm{H}}$ and 
under our simple assumptions the \trot\
values are independent of $n_{\mathrm{H}}$.
As an example of this approach, Fig.~\ref{h3o+form}b shows the \trot\
value between the $K=10$ and $K=5$ metastable levels as a function of
$\chi_{\mathrm{e}}$. We adopt here $T_{\mathrm{gas}}=50$ and $200$ K, much
lower than 500 K. Since both the recombination rate increases, and
$k_{K,K'}^{\mathrm{col}}$ is expected to decrease with decreasing \tgas, we
may expect the effect of formation pumping and \trot\ to be the highest for
the {\em lowest} \tgas. For $T_{\mathrm{gas}}=50$ K, \trot\ is much higher 
than \tgas\ even for relatively low $\chi_{\mathrm{e}}\sim10^{-6}$. 
$T_{\mathrm{rot}}\sim500$ K is then obtained for
$\chi_{\mathrm{e}}\sim10^{-4}$, typical of translucent clouds, provided that  
$k_{K,K'}^{\mathrm{col}}\lesssim10^{-10}$ cm$^3$ s$^{-1}$.
Higher $T_{\mathrm{gas}}=200$ K implies a lower recombination rate and a
presumably higher $k_{K,K'}^{\mathrm{col}}$, with the consequence that
$T_{\mathrm{rot}}$ decreases for the same $\chi_{\mathrm{e}}$. From the
excitation point of view, we thus roughly expect two {\em opposite} solutions
to explain $T_{\mathrm{rot}}\gtrsim500$ K as observed in Arp~220: low \tgas\
($\lesssim100$ K) and, of course, high $T_{\mathrm{gas}}\gtrsim500$ K.

\end{appendix}

\begin{appendix}
\section{The formation and abundance of \ohp}
\label{appc}

   \begin{figure}
   \centering
   \includegraphics[width=8.5cm]{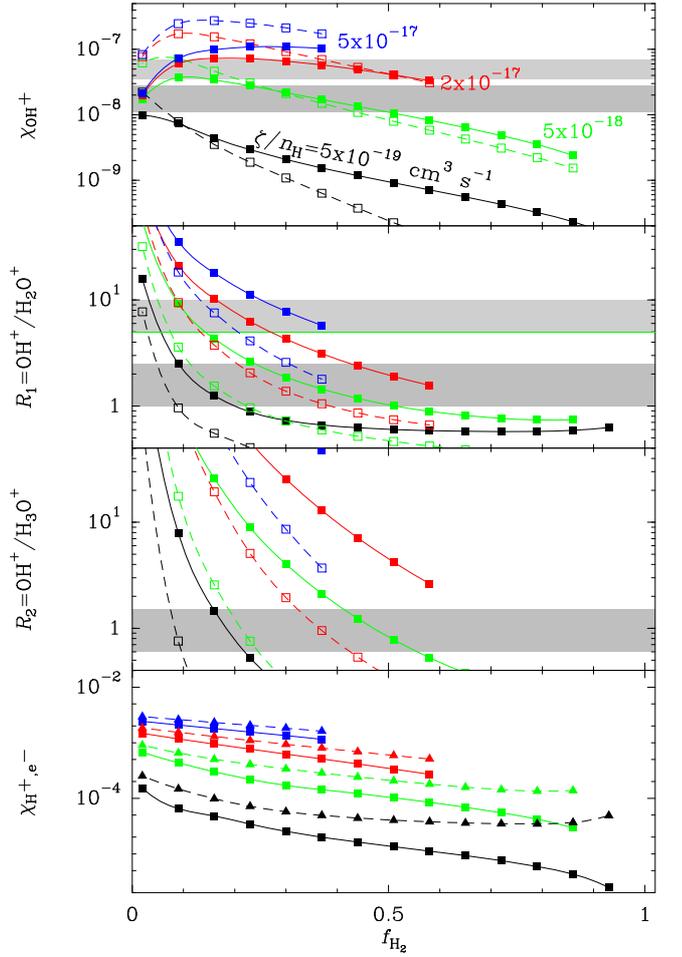}
   \caption{Predicted \ohp\ abundance (upper panel), \ohp/\hdop\ and
     \ohp/\htop\ ratios (mid panels), and H$^+$ and e$^-$ abundances (lower
     panels), as a function of the molecular fraction $f_{\mathrm{H_2}}$ (up
     to approximately $f_{\mathrm{H_2}}^{\mathrm{max}}$, see
     eq.~(\ref{fh2max})). Each curve corresponds to a value of
     $\zeta/n_{\mathrm{H}}$, indicated in the upper 
     panel in units of cm$^3$ s$^{-1}$. Solid and dashed curves correspond to
     $T_{\mathrm{gas}}=150$ and 550 K, respectively. The inferred values of
     $\chi_{\mathrm{OH^+}}$, $R_1$, and $R_2$, 
     in the nuclear regions of NGC 4418 and Arp 220, and in the \cext\
     of Arp 220, are indicated with the shaded dark and light regions,
     respectively. In the lower panel, squares and triangles show the
     predicted $\chi_{\mathrm{H^+}}$ and $\chi_{\mathrm{e}}$.}  
    \label{chemres}
    \end{figure}

Here we present the full set of calculated abundances of \ohp, \hdop,
  and \htop, together with the corresponding abundances of electrons and H$^+$,
  based on the methods discussed in Sect.~\ref{sec:chem}.  Some analytical
  expressions for \ohp\ production rates are compared, showing that the most
  likely path for \ohp\ formation is $\mathrm{O^+ + H_2}$, with the
  contribution from the charge exchange reaction $\mathrm{H^+ + OH}$ and 
  from direct OH ionization being of secondary importance even in the nuclear
  regions, where high OH abundances are inferred (Paper I). We also show that
  H$^+$ is most likely formed through X/cosmic ray ionization, rather than via
  the $\mathrm{C^+ + OH}\rightarrow\mathrm{CO^+ + H}\rightarrow\mathrm{CO +
    H^+}$ FUV/chemical path.  

The predicted \ohp\ abundance relative to H nuclei, $\chi_{\mathrm{OH^+}}$,
and both $R_1$ and $R_2$ are shown in Fig.~\ref{chemres} as a function of
$f_{\mathrm{H_2}}$ for several values of $\zeta/n_{\mathrm{H}}$, together with
the values inferred in the nuclear regions of NGC 4418 and Arp 220 (shaded
dark regions) and in the \cext\ of Arp 220 (shaded light regions). The
abundances of H$^+$ and e$^-$ are also shown in the lower panel.

The values relevant to the nuclear regions,
  $\chi_{\mathrm{OH^+}}=(1-3)\times10^{-8}$, $R_1\sim 1-2.5$, and $ R_2 \sim
  1$, can be obtained for very different values of $\zeta/n_{\mathrm{H}}$,
  provided that $f_{\mathrm{H_2}}$ attains the appropriate regime. For example,
  values compatible with observations are found for
  $\zeta/n_{\mathrm{H}}\approx5\times10^{-19}$ cm$^3$ s$^{-1}$,
  $f_{\mathrm{H_2}}\approx0.1$, and high \tgas. For low values of
  $\zeta/n_{\mathrm{H}}$, however, the range of $f_{\mathrm{H_2}}$ compatible with
  observations is very narrow. This range broadens for higher
  $\zeta/n_{\mathrm{H}}$ and $f_{\mathrm{H_2}}$, which we favor owing to the
  inferred high column densities.

According to the equilibrium $\chi_{\mathrm{H^+}}$ and $\chi_{\mathrm{e}}$, we
can compare the \ohp\ volume formation rates due to $\mathrm{O^+ + H_2}$,
$\Gamma_{\mathrm{OH^+;O^+|H_2}}$, and due to $\mathrm{O + H_3^+}$,
$\Gamma_{\mathrm{OH^+;H_3^+|O}}$, at high $f_{\mathrm{H_2}}$ \citep[see][for
diffuse clouds]{lis07}. In what follows, $k_{\mathrm{X|Y}}$ is the rate
  coefficient for the $\mathrm{X+Y}$ reaction. On the one hand, 
$\Gamma_{\mathrm{OH^+;O^+|H_2}}= 0.5 f_{\mathrm{H_2}} n_{\mathrm{H}}^2
\chi_{\mathrm{O^+}} k_{\mathrm{O^+|H_2}}$, where 
$\chi_{\mathrm{O^+}}\approx k_{\mathrm{H^+|O}} \chi_{\mathrm{O^0}}\chi_{\mathrm{H^+}} 
(0.5 f_{\mathrm{H_2}}k_{\mathrm{O^+|H_2}}+
(1-f_{\mathrm{H_2}})k_{\mathrm{O^+|H}})^{-1}$. 
Using the UMIST06 reactions rates given in Table~\ref{tab:reactions} at
$T_g=150$ K,
\begin{eqnarray}
\Gamma_{\mathrm{OH^+;O^+|H_2}} & \approx & 2.5\times10^{-18} \times
n_{\mathrm{H}}^2 \times \frac{1}{1+0.55(1-f_{\mathrm{H_2}})/f_{\mathrm{H_2}}} 
\nonumber \\ & \times & 
\frac{\chi_{\mathrm{O^0}}}{3\times10^{-4}} \times 
\frac{\chi_{\mathrm{H^+}}}{6\times10^{-5}}
\, \mathrm{cm^{-3} s^{-1}}
\label{formohplus1}
\end{eqnarray}

On the other hand, 
$\Gamma_{\mathrm{OH^+;H_3^+|O}}= n_{\mathrm{H}}^2 \chi_{\mathrm{O^0}}
k_{\mathrm{H_3^+|O}} \chi_{\mathrm{H_3^+}}$, where 
$\chi_{\mathrm{H_3^+}}\approx k_{\mathrm{H_3^+|e}}^{-1}
(f_{\mathrm{H_2}}/\chi_{\mathrm{e}}) (\zeta/n_{\mathrm{H}})$. Therefore,
\begin{eqnarray}
\Gamma_{\mathrm{OH^+;H_3^+|O}} & \approx &1.7\times10^{-19} \times n_{\mathrm{H}}^2
\times\frac{\chi_{\mathrm{O^0}}}{3\times10^{-4}} \nonumber \\ & \times &
\frac{f_{\mathrm{H_2}}/\chi_{\mathrm{e}}}{3.3\times10^3} \times
\frac{\zeta/n_{\mathrm{H}}}{2\times10^{-17}\,\mathrm{cm^3\,s^{-1}}}
\, \mathrm{cm^{-3} s^{-1}}
\label{formohplus2}
\end{eqnarray}
Taking the limit $f_{\mathrm{H_2}}\approx1$ in eq.~(\ref{formohplus1}), one
obtains 
\begin{equation}  
\frac{\Gamma_{\mathrm{OH^+;O^+|H_2}}}{\Gamma_{\mathrm{OH^+;H_3^+|O}}}  \sim 
15 \times \frac{\chi_{\mathrm{e}}/f_{\mathrm{H_2}}}{3\times10^{-4}} \times
\frac{\chi_{\mathrm{H^+}}}{6\times10^{-5}} \times
\frac{2\times10^{-17}\,\mathrm{cm^3\,s^{-1}}}{\zeta/n_{\mathrm{H}}}.
\label{formohplusratio}
\end{equation}  
The reference values in
eqs.~(\ref{formohplus1})-(\ref{formohplusratio}) are based on
$\zeta/n_{\mathrm{H}}=2\times10^{-17}$ 
cm$^3$ s$^{-1}$ at $T_g=150$ K, and indicate 
$\Gamma_{\mathrm{OH^+;O^+|H_2}}\gg\Gamma_{\mathrm{OH^+;H_3^+|O}}$;  
increasing $\zeta/n_{\mathrm{H}}$ would increase both $\chi_{\mathrm{H^+}}$ and
$\chi_{\mathrm{e}}$, thus further increasing the above ratio. 
The $\mathrm{O + H_3^+}$ mechanism only becomes competitive for 
low $\zeta/n_{\mathrm{H}}\lesssim5\times10^{-19}$ cm$^3$ s$^{-1}$ and
$f_{\mathrm{H_2}}\approx1$, but the resulting \ohp\ abundance is then
much lower than the inferred one (Fig.~\ref{chemres})
and the \ohp/\hdop\ ratio becomes lower than 1 \citep{hol12}.

   \begin{table}
      \caption[]{Relevant reaction rates involved in the \ohp, \hdop, and
        \htop\ formation and destruction. Rate coefficients are taken from the
        UMIST06 database \citep{woo07}.} 
         \label{tab:reactions}
          \begin{tabular}{ll}   
            \hline
            \noalign{\smallskip}
Reaction & Rate coefficient   \\   
         & (cm$^3$ s$^{-1}$)      \\
            \noalign{\smallskip}
            \hline
            \noalign{\smallskip}
$\mathrm{H^++O\rightarrow O^++H} $ &
$k_{\mathrm{H^+|O}}=7.31\times10^{-10}T_{300}^{0.23}\exp\{-225.9/T\}$    \\ 
$\mathrm{O^++H\rightarrow O+H^+} $    &
$k_{\mathrm{O^+|H}}=5.66\times10^{-10}T_{300}^{0.36}\exp\{8.6/T\}$    \\ 
$\mathrm{O^++H_2\rightarrow OH^++H} $ &
$k_{\mathrm{O^+|H_2}}=1.7\times10^{-9}$  \\ 
$\mathrm{OH^++H_2\rightarrow H_2O^++H} $ &
$k_{\mathrm{OH^+|H_2}}=1.01\times10^{-9}$   \\  
$\mathrm{H_2O^++H_2\rightarrow H_3O^++H} $ &
$k_{\mathrm{H_2O^+|H_2}}=6.4\times10^{-10}$   \\  
$\mathrm{H^++OH\rightarrow OH^++H} $ &
$k_{\mathrm{H^+|OH}}=2.1\times10^{-9}$    \\ 
$\mathrm{H^++H_2O\rightarrow H_2O^++H} $ &
$k_{\mathrm{H^+|H_2O}}=6.9\times10^{-9}$    \\ 
$\mathrm{H_2^++H_2\rightarrow H_3^++H} $ &
$k_{\mathrm{H_2^+|H_2}}=2.08\times10^{-9}$    \\ 
$\mathrm{H_3^++O\rightarrow OH^++H_2} $ &
$k_{\mathrm{H_3^+|O:OH^+}}=8.4\times10^{-10}$    \\ 
$\mathrm{H_3^++O\rightarrow H_2O^++H} $ &
$k_{\mathrm{H_3^+|O:H_2O^+}}=3.6\times10^{-10}$    \\ 
$\mathrm{H_3^++OH\rightarrow H_2O^++H_2} $ &
$k_{\mathrm{H_3^+|OH}}=1.3\times10^{-9}$    \\ 
$\mathrm{H_3^++H_2O\rightarrow H_3O^++H_2} $ &
$k_{\mathrm{H_3^+|H_2O}}=5.9\times10^{-9}$    \\ 
$\mathrm{OH^++e\rightarrow O+H} $ &
$k_{\mathrm{OH^+|e}}=3.75\times10^{-8}T_{300}^{-0.5}$   \\  
$\mathrm{H_2O^++e\rightarrow products} $ &
$k_{\mathrm{H_2O^+|e}}=4.3\times10^{-7}T_{300}^{-0.5}$   \\  
$\mathrm{H_3O^++e\rightarrow products} $ &
$k_{\mathrm{H_3O^+|e}}=4.3\times10^{-7}T_{300}^{-0.5}$   \\  
$\mathrm{H_3^++e\rightarrow products} $ &
$k_{\mathrm{H_3^+|e}}=6.7\times10^{-8}T_{300}^{-0.52}$    \\ 
               \noalign{\smallskip}
            \hline
         \end{tabular} 
\begin{list}{}{}
\item[$^{\mathrm{a}}$] $T_{300}$ is $T/\mathrm{300 \,K}$; $T$ denotes \tgas.  
\end{list}
   \end{table}

In the nuclear regions of both NGC 4418 and Arp 220, high reservoirs of
OH and \hdo\ are found, with estimated abundances 
$\chi_{\mathrm{H_2O}}\sim (0.5-1)\times10^{-5}$ and
$\mathrm{OH/H_2O}\sim0.5$ (Paper I). 
The \ohp\ volume formation rate due to $\mathrm{H^++OH}$ is given by
$\Gamma_{\mathrm{OH^+;H^+|OH}}= n_{\mathrm{H}}^2
\chi_{\mathrm{H^+}}\chi_{\mathrm{OH}} k_{\mathrm{H^+|OH}}$, and so
\begin{eqnarray}
\Gamma_{\mathrm{OH^+;H^+|OH}}\approx 3.8\times10^{-19} n_{\mathrm{H}}^2
\times \frac{\chi_{\mathrm{OH}}}{3\times10^{-6}} \times 
\frac{\chi_{\mathrm{H^+}}}{6\times10^{-5}}
\, \mathrm{cm^{-3} s^{-1}},
\end{eqnarray}
which is significantly lower than 
$\Gamma_{\mathrm{OH^+;O^+|H_2}}$ in eq.~(\ref{formohplus1}) if there is
enough free oxygen. However, a significant fraction of O$^0$ in these
high density regions is expected to be locked into CO, \hdo, OH and O$_2$.
Depletion of free O$^0$ into molecules in regions dominated by hot core
chemistry (Paper I) would thus reduce the \ohp\ formation rate in 
eq.~(\ref{formohplus1}), thus increasing the potential relative importance of
$\Gamma_{\mathrm{OH^+;H^+|OH}}$. There is, however, an observational drawback
for the potential importance of $\mathrm{H^++OH}$. 
The rate of \hdop\ formation due to $\mathrm{H^++H_2O}$ is higher than
$\Gamma_{\mathrm{OH^+;H^+|OH}}$, because both
$\chi_{\mathrm{H_2O}}>\chi_{\mathrm{OH}}$ (Paper I) and 
$k_{\mathrm{H^+|H_2O}}>k_{\mathrm{H^+|OH}}$ (Table~\ref{tab:reactions}).
Thus, if the \ohp\ were formed via $\mathrm{H^++OH}$, one would also expect
an additional, important contribution to \hdop\ formation due to
$\mathrm{H^++H_2O}$, and then $R_1$ would fall below the inferred value of
$1-2.5$. Likewise, direct ionization of OH due to UV photons is expected to be
of secondary importance, because \hdo\ has a higher cross-section for
ionization. As we argue in \S\ref{sec:zeta}, the observed lines of molecular
ions are most probably formed in regions where the formation rate is enhanced
due to the presence of high quantities of atomic oxygen, and where the ions
survive sufficiently long due to the presence of moderate densities.

According to the above discussion, model results for $\chi_{\mathrm{OH^+}}$
and $R_1$ relevant for the current observations can be basically
understood in terms of the  
$\mathrm{H^+ \rightarrow O^+\rightarrow OH^+\rightarrow H_2O^+\rightarrow
  H_3O^+}$ sequence, but how is the H$^+$ produced? 
We have been assuming in the analysis that
the H$^+$ ionization is caused by X/cosmic rays, but the possibility of
enhanced H$^+$ production by chemical routes, as discussed by \cite{ste95} and
\cite{hol12}, is worth considering. According to their models for high density
PDRs illuminated by strong FUV fields, in the hot transition region between
the H {\sc i} and H$_2$ zones, a major source of H$^+$ formation 
is given by $\mathrm{OH+C^+ \rightarrow CO^++ H}$ followed by 
$\mathrm{CO^++ H \rightarrow CO + H^+}$. OH would be produced via the
endothermic $\mathrm{O+ H_2 \rightarrow OH + H}$, that is enhanced provided
that a fraction of H$_2$ is vibrationally excited. The volume rate of H$^+$
production is given by
$\Gamma_{\mathrm{H^+;OH|C^+}} = n_{\mathrm{H}}^2 k_{\mathrm{OH|C^+}} \chi_{\mathrm{OH}}
 \chi_{\mathrm{C^+}} (1+\alpha)^{-1}$,
where $\alpha$ stands for the chance that CO$^+$ reacts with H$_2$ rather than
with H {\sc i}. In the limit $\alpha=0$, and using the value for
$k_{\mathrm{OH|C^+}}$ given by \cite{hol12},
\begin{equation}
\Gamma_{\mathrm{H^+;OH|C^+}} = 2\times10^{-18} n_{\mathrm{H}}^2  
\times \frac{\chi_{\mathrm{C^+}}}{1.4\times10^{-4}} \times 
\frac{\chi_{\mathrm{OH}}}{5\times10^{-6}}
\, \mathrm{cm^{-3} s^{-1}}
\end{equation}
for $T_{\mathrm{gas}}=300$ K; the reference value for $\chi_{\mathrm{OH}}$ is
taken from Paper I. This is equivalent to a cosmic ray ionization
per unit density of $\zeta/n_{\mathrm{H}}\sim2\times10^{-18}$ cm$^3$ s$^{-1}$,
and predicts a peak of \ohp\ abundance of $\sim10^{-8}$ as well as $R_1$ and
$R_2$ values of $\sim1$ \citep{ste95,hol12}, as roughly inferred in both
NGC~4418 and Arp~220. 
This FUV/chemical route for the production of H$^+$ has in the
present case, however, one main drawback: the nuclear \ohp\
is observed over a range in visual extinction of $\Delta A_V\sim200$ mag,
while theoretical predictions indicate a peak thickness in the \ohp\ abundance
of $\Delta A_V<1$ mag where C$^+$ is still abundant and \tgas\ remains  
high. It seems very unlikely that FUV radiation maintains basically all
carbon ionized, and a relatively low H$_2$ fraction, over columns of $\Delta
A_V\sim200$ mag along the line of sight and over spatial scales of $20-30$ and 
$80-150$ pc in NGC~4418 and Arp~220, 
respectively \citep[see also models by][]{abe09}. 
Even if $10^6$ O stars were packed into a region of $\sim10$ pc radius
(NGC~4418), the column density between neighboring stars would be $\sim10^{23}$
cm$^{-2}$, and the quickly absorbed FUV radiation would only affect relatively
compact regions around the stars.
In summary, the inferred columns of O-bearing
cations favor a source of ionization that can penetrate through large columns
of gas, indicating the importance of X and cosmic rays. 

The dependence of $\chi_{\mathrm{OH^+}}$ on both $f_{\mathrm{H_2}}$ and
$\zeta/n_{\mathrm{H}}$ in Fig.~\ref{chemres} can be understood as follows.
For given $f_{\mathrm{H_2}}$, $\chi_{\mathrm{OH^+}}$
scales linearly with $\zeta/n_{\mathrm{H}}$ for relatively
low values of $\zeta/n_{\mathrm{H}}$, but {\em saturates} for sufficiently
high values of $\zeta/n_{\mathrm{H}}$. Saturation occurs when the destruction 
of \ohp\ is dominated by recombination, rather than by reaction with H$_2$
(producing \hdop). In the saturated regime, further increase of
$\zeta/n_{\mathrm{H}}$ increases by the same factor the \ohp\ formation
rate and the \ohp\ destruction rate per unit of \ohp\ molecule, leaving
$\chi_{\mathrm{OH^+}}$ unchanged. The condition for saturation is given by 
$\chi_{\mathrm{e}}k_{\mathrm{OH^+|e}}\gg0.5f_{\mathrm{H_2}}k_{\mathrm{OH^+|H_2}}$,
where $k_{\mathrm{OH^+|e}}\approx5\times10^{-8}$ cm$^3$ s$^{-1}$ at 150 K, and
$k_{\mathrm{OH^+|H_2}}=10^{-9}$ cm$^3$ s$^{-1}$. The above condition
yields $\chi_{\mathrm{e}}\gg10^{-2}\times f_{\mathrm{H_2}}$. Once saturated,
$\chi_{\mathrm{OH^+}}$ can be approximated analytically by
\begin{eqnarray}
\chi_{\mathrm{OH^+}}^{\mathrm{sat}} & \approx & 
\frac{k_{\mathrm{H^+|O}}}{k_{\mathrm{OH^+|e}}} \times
\frac{0.5f_{\mathrm{H_2}}k_{\mathrm{O^+|H_2}}}{0.5f_{\mathrm{H_2}}k_{\mathrm{O^+|H_2}}
+(1-f_{\mathrm{H_2}})k_{\mathrm{O^+|H}}} \times \chi_{\mathrm{O^0}} \nonumber
\\
& \approx &
7.9\times10^{-7}\times \frac{1}{1+0.55(1-f_{\mathrm{H_2}})/f_{\mathrm{H_2}}} 
\times\frac{\chi_{\mathrm{O^0}}}{3\times10^{-4}}
\label{eq:satur}
\end{eqnarray}
giving, for $T_{\mathrm{gas}}=150$ K,
$\chi_{\mathrm{OH^+}}^{\mathrm{sat}}\approx2.7\times10^{-8}$ and 
$3.4\times10^{-7}$ for $f_{\mathrm{H_2}}=0.02$ and $0.3$, respectively, in
rough agreement with the values in Fig.~\ref{chemres}.

The dependence of $\chi_{\mathrm{OH^+}}$ on $f_{\mathrm{H_2}}$ is
different in the saturated and unsaturated regimes. According to
eq.~(\ref{eq:satur}), $\chi_{\mathrm{OH^+}}$ increases with $f_{\mathrm{H_2}}$
in the saturated regime, because the $\mathrm{O^++H_2}$ reaction producing
\ohp\ is increasingly favored relative to $\mathrm{O^++H}$, which
transfers the charge back to atomic hydrogen. However, in the unsaturated
regime $\chi_{\mathrm{OH^+}}$ {\em decreases} with increasing
$f_{\mathrm{H_2}}$ (Fig.~\ref{chemres}). In this case, \ohp\ destruction is
dominated by $\mathrm{OH^++H_2}$ and $\chi_{\mathrm{OH^+}}$ is thus 
analitically aproximated as 
\begin{eqnarray}
\chi_{\mathrm{OH^+}}^{\mathrm{unsat}} & \approx & 
\frac{k_{\mathrm{H^+|O}}}{k_{\mathrm{OH^+|H_2}}} \times
\frac{k_{\mathrm{O^+|H_2}}\,\chi_{\mathrm{H^+}}}
{0.5f_{\mathrm{H_2}}k_{\mathrm{O^+|H_2}} 
+(1-f_{\mathrm{H_2}})k_{\mathrm{O^+|H}}} \times \chi_{\mathrm{O^0}} \nonumber
\\
& \approx &
1.5\times10^{-8}\times \frac{1}{1+0.82f_{\mathrm{H_2}}} 
\times\frac{\chi_{\mathrm{O^0}}}{3\times10^{-4}} \times 
\frac{\chi_{\mathrm{H^+}}}{10^{-4}}
\label{eq:unsatur}
\end{eqnarray}
Therefore, $\chi_{\mathrm{OH^+}}^{\mathrm{unsat}}$ decreases with increasing
$f_{\mathrm{H_2}}$. For given $\zeta/n_{\mathrm{H}}$ (i.e. for each curve in
Fig.~\ref{chemres}), a maximum of \ohp\ is reached between the saturated
regime (low $f_{\mathrm{H_2}}$) and the unsaturated one (high $f_{\mathrm{H_2}}$).

\end{appendix}

\end{document}